%% file: ifacconf.tex
\documentclass{ifacconf}

\usepackage{graphicx}      
\usepackage{natbib}        
\usepackage{amssymb}  
\usepackage{amsfonts} 
\usepackage{epstopdf}
\usepackage{color}
\usepackage{bbding}

\usepackage{algorithm}  
\usepackage{algpseudocode}  
\usepackage{amsmath}  

\input{def}
\begin{document}
\begin{frontmatter}

\title{Online Observability of \\ 
	Boolean Control Networks\thanksref{footnoteinfo}} 

\thanks[footnoteinfo]{This paper is funded by the Special Foundation for Basic Science and Frontier Technology Research Program of Chongqing with No.cstc2017jcyjAX0295, the Capacity Development Grant of Southwest University with No.SWU116007, the National Science Foundation of China under Grant No. 61802318 and Fundamental Research Funds for the Central Universities SWU118014}

\author[First]{Guisen~Wu}, 
\author[First]{Liyun~Dai\Envelope}, 
\author[First]{Zhiming~Liu},
\author[Second]{Taolue~Chen},
\author[Third]{Jun~Pang}
\address[First]{The School of Computer and Information Science, Southwest University
Chongqing, 400715 China. (e-mail: \{wgs233,~dailiyun,~zhimingliu88\}@swu.edu.cn)}
\address[Second]{The Department of Computer Science and Information Systems, Birkbeck, University of London, (e-mail: taolue@dcs.bbk.ac.uk)}
\address[Third]{The Faculty of Science, Technology and Communication
and the Interdisciplinary Centre for Security, Reliability and Trust, University of Luxembourg, (e-mail: jun.pang@uni.lu)}

\maketitle
\begin{abstract}                
Observabililty is an important topic of Boolean control networks (BCNs). In this paper, we propose a new type of observability named online observability to present the sufficient and necessary condition of determining the initial states of BCNs, when their initial states cannot be reset. And we design an algorithm to decide whether a BCN has the online observability. Moreover, we prove that a BCN is identifiable iff it satisfies controllability and the online observability, which reveals the essence of identification problem of BCNs.
\end{abstract}

\begin{keyword}
Boolean control network, Observability, Controllability, Identification, Algorithm 
\end{keyword}

\end{frontmatter}

\input{sec/introduction}

\input{sec/preliminaries}

\input{sec/onlineobservability}

\input{sec/determining}
\input{sec/identification}
\input{sec/conclusions}

\bibliography{ifacconf}             
                                                   







\appendix
\input{sec/appendix}

\end{document}

%% file: def.tex
\newtheorem{example}{Example}

\newtheorem{definition}{Definition}

\newtheorem{lemma}{Lemma}

\def \BN {{\em BN}}
\def \BNs {{\em BNs}}
\def \BCN {{\em BCN}}
\def \BCNs {{\em BCNs}}

\def \Ks {\Gamma}
\def \Ri {\psi}

\def \Ded {\zeta}
\def \BB {{\mathcal{B}}}



%% file: sec/introduction.tex

\section{Introduction}
\label{sec:intro}

In 1960s, Nobel Prize laureates Jacob and Monod claimed ``Any cell contains a number of regulatory genes that act as switches and can turn one another on and off (\cite{Jacob1961Genetic}). If genes can turn one another on and off, then you can have genetic circuits.'' Inspired by these Boolean-type actions in genetic circuits, Boolean networks (\BNs) were proposed by Kauffman for modeling nonlinear and complex biological systems (\cite{Kauffman1968Metabolic}). 


\subsection{Boolean control networks}
A natural extension of \BN\ is  Boolean control network (\BCN) with external regulations and perturbations (\cite{Ideker2001A}). \BCNs\ have been applied to  various real-life problems 
and typical examples including  structural and functional analysis of signaling and regulatory networks (\cite{Kaufman1999A} and \cite{Klamt2006A}), 
abduction based drug target discovery (\cite{Biane2017Abduction}), 
and pursuing evasion problems in polygonal environments (\cite{Thunberg2011A}).

A \BCN\ has three distinct finite sets of nodes $\{\mathsf{i}_1,\ldots, \mathsf{i}_\ell\}$, $\{\mathsf{s}_1,\ldots,\mathsf{s}_m\}$ and $\{\mathsf{o}_1,\ldots, \mathsf{o}_n\}$ for some natural numbers $\ell$, $m$ and $n$, which are called the {\em input-nodes}, {\em state-nodes}  and {\em output-nodes}, respectively. As a {\BCN} $\BB$, each node of   $\BB$   takes a Boolean value at any time point. We use $\mathsf{i}(t)$, $\mathsf{s}(t)$ and $\mathsf{o}(t)$  to denote the vectors of values $(\mathsf{i}_1(t),\ldots,\mathsf{i}_\ell(t))$ of the input-nodes,  $(\mathsf{s}_1(t),\ldots,\mathsf{s}_m(t))$ of the state-nodes, and $(\mathsf{o}_1,\ldots,\mathsf{o}_n(t))$ of the output-nodes  which are called the {\em input}, {\em state} and {\em output} at time $t$, respectively.
\begin{itemize}
\item The state $\mathsf{s}(t+1)$ at time $t+1$ is determined by the input $\mathsf{i}(t)$ and state $\mathsf{s}(t)$ at time point $t$, that is there is a Boolean functions $\sigma$ such that $\mathsf{s}(t+1)=\sigma(\mathsf{i}(t),\mathsf{s}(t))$, 
	\item and the output $\mathsf{o}(t)$ at time $t$ is determined by the state $\mathsf{s}(t)$ at time $t$, that is there exists a Boolean functions $\rho$ such that $\mathsf{o}(t)=\rho(\mathsf{s}(t))$.
	
\end{itemize} 

These two functions are called the {\em updating rules} of $\BB$. And a {\em timed run}  (or {\em execution}) up to any time point $t$ of $\BB$ is a sequence $\mathsf{I}[t]=\mathsf{i}(0)\ldots \mathsf{i}(t)$ of inputs, a sequence $\mathsf{S}[t]=\mathsf{s}(0)\ldots \mathsf{s}(t)$ of states, and a sequence $\mathsf{O}[t]=\mathsf{o}(0)\ldots \mathsf{o}(t)$  of outputs. The sequences $\mathsf{S}[t]$ and $\mathsf{O}[t]$ are produced by the initial state $\mathsf{s}(0)$ and the input sequence $\mathsf{i}(0)\ldots \mathsf{i}(t-1)$.

\subsection{Related work: control-theoretical properties}
 In general, the updating rules are complex and it is difficult to analyze dynamic properties of a {\BCN} $\BB$, such as if it is possible and how to  find a sequence $\mathsf{I}[t]$ of inputs so that a state $\mathsf{s}'$ is reachable from an initial state $\mathsf{s}$,  and whether it is possible and how to find input sequences so that  they and their corresponding output sequences can  determine the initial state.  These are regarded as control-theoretical issues of {\em controllability, observability, reconstructibility} and {\em identifiability} in the study of \BCNs\ as  control theory of dynamic systems (\cite{Akutsu2007Control,cheng2009controllability, Zhao2010Input, Cheng2011Identification, Cheng2011Analysis,Fornasini2013Observability,Pangjun,Pangjuna} and \cite{Zhang2015A}).
We summarize  the notions of these control-theoretical properties as follows.

\smallskip\noindent
{\em Controllability.}  The controllability was proposed in~\cite{Akutsu2007Control} which 
 states that a \BCN\ $\BB$ is {\em controllable} 
if for any pair of states $\mathsf{s}$ and $\mathsf{s}'$, there  exists an input sequence $\mathsf{I}[t]=\mathsf{i}(0)\ldots\mathsf{i}(t)$ for some $t>0$ such that $\mathsf{s}(0)\ldots\mathsf{s}(t+1)$ is the corresponding sequence of states with $\mathsf{s}(0) =\mathsf{s}$ and $\mathsf{s}(t+1)=\mathsf{s}'$. And the  problem of the controllability of \BCNs\ is known to be {NP}-hard (\cite{Akutsu2007Control}). 
%

\smallskip\noindent
{\em Observability.} This is in general about if it is possible and how to determine the initial state 
a \BCN\   by providing sequences of inputs and observing the corresponding output sequences, when the updating rules are available (\cite{cheng2009controllability}). Now four types of observability have been proposed in the literature (\cite{Zhang2016Observability}).


\begin{enumerate}
\item The  {\sf Type-I}  observability was proposed in~\cite{cheng2009controllability}. It states that a {\BCN} is {\em observable} if for any initial state $\mathsf{s}(0)$, there is an input sequence $\mathsf{I}[t]=\mathsf{i}(0)\ldots\mathsf{i}(t)$  such that  for any  $\mathsf{s}'(0)$ different from $\mathsf{s}(0)$, the output sequence  $\mathsf{o}(0)\ldots\mathsf{o}(t+1)$ produced by  $\mathsf{I}[t]$ from $\mathsf{s}(0)$ is different from that   $\mathsf{o}'(0)\ldots \mathsf{o}'(t+1)$ produced by $\mathsf{I}[t]$ from $\mathsf{s}'(0)$. 

	
\item The  {\sf Type-II} observability was proposed in \cite{Zhao2010Input}. It states that a \BCN\  is {\em observable} if for any two different initial states $\mathsf{s}(0)$ and $\mathsf{s}'(0)$, there exists an input sequence  $\mathsf{I}[t]=\mathsf{i}(0)\ldots\mathsf{i}(t)$ for some $t>0$ such that the output sequences $\mathsf{o}(0)\ldots\mathsf{o}(t+1)$ and  $\mathsf{o}'(0)\ldots\mathsf{o}'(t+1)$ corresponding to  the different initial states  $\mathsf{s}(0)$ and $\mathsf{s}'(0)$, respectively, are different.

\item The  {\sf Type-III} observability was proposed in~\cite{Cheng2011Identification} and it states that a \BCN\  is observable if there exists an input sequence  $\mathsf{I}[t]=\mathsf{i}(0)\ldots\mathsf{i}(t)$ for some $t>0$ such that for any two different initial states $\mathsf{s}(0)$ and $\mathsf{s}'(0)$, the output sequence $\mathsf{o}(0)\ldots\mathsf{o}(t+1)$  corresponding to $\mathsf{s}(0)$  is different from that  $\mathsf{o}'(0)\ldots\mathsf{o}'(t+1)$ corresponding to $\mathsf{s}'(0)$.



	
\item The  {\sf Type-IV}  observability was  proposed in~\cite{Fornasini2013Observability} which states that a \BCN\ is observable if there is a  natural number $N$ such that  the output sequences $\mathsf{o}(0) \ldots\mathsf{o}(t+1)$ and  $\mathsf{o}'(0)\ldots\mathsf{o}'(t+1)$  generated from any two different initial states $\mathsf{s}(0)$ and $\mathsf{s}'(0)$ by any input sequence $\mathsf{I}[t]=\mathsf{i}(0)\ldots\mathsf{i}(t)$ are different if $t\geq N$.

\end{enumerate}
\smallskip\noindent
{\em Reconstructibility.} This is in general about if it is possible and how to 
determine the current state 
a \BCN\   by its iuput and output sequence, when we know the updating rules. There are two types of reconstructibility have been proposed in the literature (\cite{Kuize2020Detectability}).
\begin{enumerate}
\item The  {\sf Type-I} reconstructibility was proposed in~\cite{Fornasini2013Observability} which states that a \BCN\ is reconstructible if there is a  natural number $N$ such that  the output sequences $\mathsf{o}(0) \ldots\mathsf{o}(t+1)$ and  $\mathsf{o}'(0)\ldots\mathsf{o}'(t+1)$  generated from any two different initial states $\mathsf{s}(0)$ and $\mathsf{s}'(0)$ by any input sequence $\mathsf{I}[t]$ are different if $\mathsf{s}(t+1)\ne\mathsf{s}'(t+1)$ and $t\geq N$.
\item The  {\sf Type-II} reconstructibility was proposed in~\cite{Zhang2015A} and it states that a \BCN\ is reconstructible if there exists an input sequence  $\mathsf{I}[t]$ for some $t>0$ such that for any two different initial states $\mathsf{s}(0)$ and $\mathsf{s}'(0)$, the output sequence $\mathsf{o}(0)\ldots\mathsf{o}(t+1)$  corresponding to $\mathsf{s}(0)$  is different from that  $\mathsf{o}'(0)\ldots\mathsf{o}'(t+1)$ corresponding to $\mathsf{s}'(0)$ if $\mathsf{s}(t+1)\ne\mathsf{s}'(t+1)$.
\end{enumerate}

As it summarized in~\cite{Kuize2020Observability}, the uses of four types of observability in determining the initial state of a {\BCN} are different. 
We can call the {\sf Type-I \& II} observability as {\em strong multiple-experiment observability} and {\em multiple-experiment observability}, respectively, as their algorithms for determining the initial state of a {\BCN} of  require that the initial state of the {\BCN} can be reset again and again to repeatedly run the system. The algorithms  of {\sf Type-III \& IV}  observability for determining  the initial state assume that the initial state of the {\BCN} cannot be reset. And  the algorithm of {\sf Type-IV}  observability requires that any sufficient long input sequence can determine the initial state. Therefore, the {\sf Type-III \& IV}  observability are called as {\em single-experiment observability} and {\em arbitrary-experiment observability}, respectively. The uses of two types of reconstructibility in determining the current state of a {\BCN} are also different (\cite{Kuize2020Detectability}). The authors call the {\sf Type-I \& II} reconstructibility as  {\em arbitrary-experiment reconstructibility} and {\em single-experiment reconstructibility}, respectively, for the same reason. And as the reconstructibility is propoesd to determine the current state of a \BCN, there are not such the notions of {\em multiple-experiment reconstructibility} and {\em strong multiple-experiment reconstructibility}.

\smallskip\noindent
{\em Identifiability.} The identifiability was proposed in~\cite{Cheng2011Identification} which states that a \BCN\ is identifiable if its updating rules $\sigma$ and $\rho$ can be uniquely obtained via a proper input-output data $\mathsf{i}(0)\ldots \mathsf{i}(t)$ and $\mathsf{o}(0)\ldots \mathsf{o}(t)$. 
 
The relationship between the identifiability and other properties is complex. 
For instance, it was proposed in~\cite{Cheng2011Identification} that a \BCN\  is identifiable if and only if it satisfies controllability and observability of {\sf Type-III}. But latter in~\cite{inproceedings}, researchers state that a \BCN\  is identifiable if and only if it is controllable and satisfies the {\sf Type-II} reconstructibility. 
\subsection{Our contribution}
In~\cite{Kuize2020Observability}, reseachers call the {\sf Type-III} observability as single-experiment observability because they regard it as the sufficient and necessary condition of determining the initial state of a \BCN\ when its initial state cannot be reset. But we find that it is not, and we propose a new type of observability named {\em online observability} to present this sufficient and necessary condition. 
The reason why we call this new type of observability as online observability will be illustrated in {\em Section \ref{sec:online}}. We design an algorithm for deciding if a \BCN\ has online observability as our second contribution. Lastly, we prove that the online observability, together with the controllability, is sufficient and necessary for the identifiability, which is different from the proposions mentioned in the previous subsection. 

\smallskip\noindent
The remainder of this paper is organized as follows.
 In {\em Section \ref{sec:pre}}, we introduce the necessary preliminaries, including the formal definition and related control-theoretical properties of a \BCN. We present the formal definition of online observability {\em Section \ref{sec:online}}. We show  the  decision  algorithm of the online observability in {\em Section \ref{sec:deter}}. We prove a \BCN\ is identifiable iff it is online observable and controllable in {\em Section \ref{sec:iden}}. We draw our conclusions in  {\em Section \ref{sec:con}}.

%% file: sec/preliminaries.tex

\section{Preliminaries} 
\label{sec:pre}
We now introduce the formal definition of a {\BCN} and its control theoretical properties. Throughout the paper, we use $\mathbb{B}$ to denote the set of Boolean values $\{0,1\}$ and $\mathbb{T}$ to denote the set of discrete time domain represented by the set of  natural numbers.

\subsection{Boolean Control Networks}

We take the definition in ~\cite{Ideker2001A} in which a Boolean control network (\BCN)  is given as a directed graph together with two  Boolean valued functions which define the updating rules for the  values of the nodes. 

\begin{figure}
		\begin{center}
			\includegraphics[width=5cm]{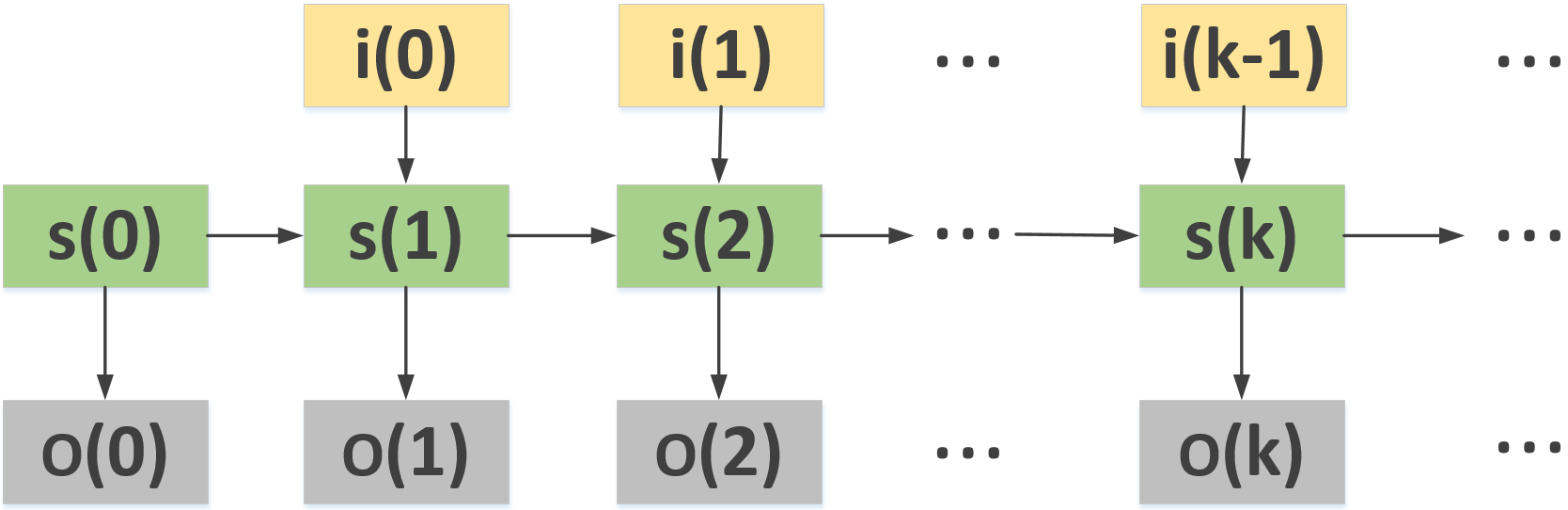}    
			\caption{The relationship between inputs, states and outputs.} 
			\label{fig:10}
		\end{center}
	\end{figure}

\begin{definition}[Boolean Control Network] 
\label{def:BCN}
A \BCN\ is a tuple $\BB = (I,S,O, E, \sigma, \rho)$, where 
\begin{itemize}
\item $I$, $S$ and $O$ are three finite nonempty disjoint sets of nodes (or vertices)
\begin{itemize}
	\item {\bf input-nodes} $I=\{\mathsf{i}_1$,\ldots ,$\mathsf{i}_{\ell}\}$,
	\item {\bf state-nodes}  $S= \{\mathsf{s}_1$,\ldots ,$\mathsf{s}_m\}$, and 
	\item {\bf output-nodes}: $O= \{\mathsf{o}_1$,\ldots ,$\mathsf{o}_n\}$.
	\end{itemize}	
	Each node is a Boolean variable which can take values in $\mathbb{B}$.
\item  $E \subseteq ((I\cup S)\times S)\cup (S\times O)$ is a set of edges among the nodes, and we say node $v$ directly affects node  $v'$  when $(v,v')$ is an edge in $E$.
\item The Boolean valued functions $\sigma: \mathbb{B}^\ell\times  \mathbb{B}^m \mapsto \mathbb{B}^m$ and $\rho: \mathbb{B}^m\mapsto \mathbb{B}^n$ are functions from the pairs of $\ell$-dimension and $m$-dimension vectors  of Boolean values to  the $m$-dimension vectors of Boolean values and from the  $m$-dimension vectors to the  $n$-dimension vectors of Boolean values, respectively. 
\item {\em Updating rules }:  We use 
 {\em input} $\mathsf{i}= (\mathsf{i}_1,\ldots, \mathsf{i}_\ell)$, {\em state} $\mathsf{s}= (\mathsf{s}_1,\ldots, \mathsf{s}_m)$ and {\em output } $\mathsf{o}= (\mathsf{o}_1,\ldots, \mathsf{o}_n)$ to denote the three Boolean vectors variables corresponding  to the input-nodes, state-nodes and output-nodes. At any time $t\in \mathbb{T}$ during the execution 
 of $\BB$, each of the variables  $\mathsf{i}$, $\mathsf{s}$ and $\mathsf{o}$ take a vector of Boolean values $\mathsf{i}(t)$, $\mathsf{s}(t)$, $\mathsf{o}(t)$  in  $\mathbb{B}^\ell$, $\mathbb{B}^m$ and $\mathbb{B}^n$, respectively, such that the following equations are satisfied.
\begin{equation}\label{equ:1}
\begin{split}
\mathsf{s}(t+1)=&\sigma(\mathsf{i}(t),\mathsf{s}(t))\\
\mathsf{o}(t)=&\rho(\mathsf{s}(t))
\end{split}
\end{equation}
The above equations are also assumed to satisfy the following two conditions
\begin{enumerate}
\item the value $\mathsf{s}_k(t+1)$ in $\mathsf{s}(t+1)$ is affected by the value  $\mathsf{i}_j(t)$ of an input node $\mathsf{i}_j \in I$ (or by the value $\mathsf{s}_j(t)$ of a state node $\mathsf{s}_j\in S$) at time $t$ only when $(\mathsf{i}_j, \mathsf{s}_k)\in E$ (or $(\mathsf{s}_j, \mathsf{s}_k)\in E$, respectively); and
\item the value $\mathsf{o}_k(t)$ in $\mathsf{o}(t)$ of an output node $\mathsf{o}_k\in O$ is affected by the value $\mathsf{s}_j(t)$ of a state node $\mathsf{s}_j\in S$ only when $(\mathsf{s}_j,\mathsf{o}_k)\in E$.
\end{enumerate}
\end{itemize}

\end{definition}
 The two updating functions are called the  {\bf updating rules}  of $\BB$. They define the values of the state-nodes at any time by the values of the input-nodes and state-nodes at the previous time point, and the values of the output-nodes by the values of the state-nodes, respectively. Such that, the relationship between inputs, states and outputs can be represented by Fig.~\ref{fig:10}. And we use 
 $\mathcal{I}_\BB$,
 $\mathcal{S}_\BB$,
 and $\mathcal{O}_\BB$ to denote  the sets of all possible inputs, states and outputs of $\BB$, respectively. 
  We will omit the subscript $\BB$ of $\mathcal{I}_\BB$, $\mathcal{S}_\BB$ and $\mathcal{O}_\BB$ when there is no confusion. 
	\begin{figure}
		\begin{center}
			\includegraphics[width=5cm]{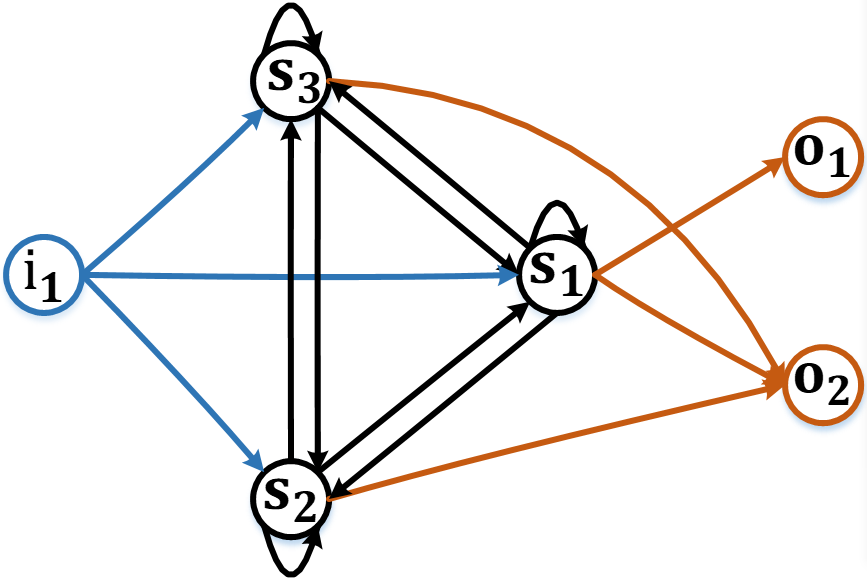}    
			\caption{A Boolean control network.} 
			\label{fig:1}
		\end{center}
	\end{figure}
	\begin{figure}
		\begin{center}
			\includegraphics[width=4cm]{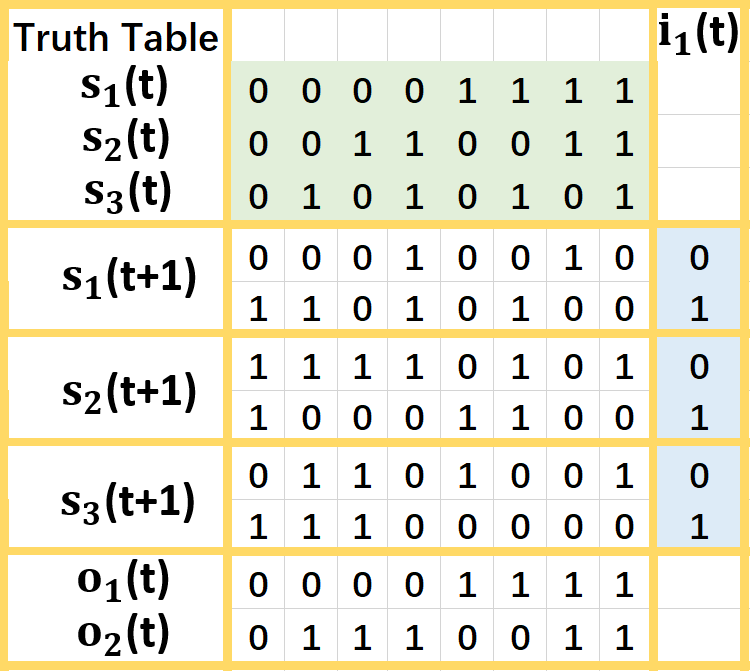}    
			\caption{The truth table which describe the updating rules of the \BCN\ shown in Fig.~\ref{fig:1}.} 
			\label{fig:2}
		\end{center}
	\end{figure}

\begin{figure}
		\begin{center}
			\includegraphics[width=5cm]{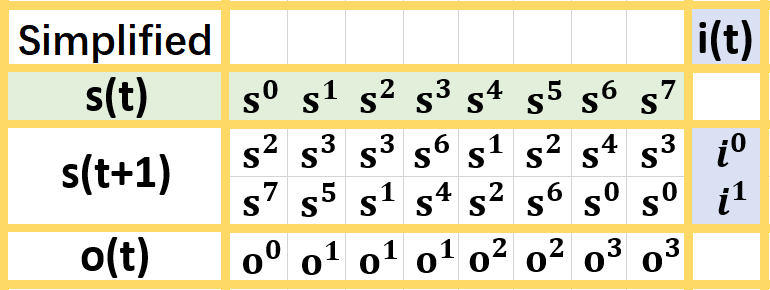}    
			\caption{The simplified  form of the updating rules.} 
			\label{fig:7}
		\end{center}
	\end{figure}

\begin{example}\label{exa:2}
Let $\BB$ be the \BCN\  shown in Fig.~\ref{fig:1} which has one  input-node $I=\{\mathsf{i}_1\}$, three state-nodes $S=\{\mathsf{s}_1, \mathsf{s}_2,\mathsf{s}_3\}$ and two output-nodes $O=\{\mathsf{o}_1,\mathsf{o}_2\}$. And 
the updating rules $\sigma: \mathbb{B}^{1}\times \mathbb{B}^{3}\mapsto \mathbb{B}^3$ and $\rho:\mathbb{B}^3\mapsto \mathbb{B}^2$ are given in the truth table Fig.~\ref{fig:2} from which the updating rules in terms of logic functions can be easily constructed.   For instance, 
the updating rule of output-node $\mathsf{o}_1$ is 
$\mathsf{o}_1(t)=\mathsf{s}_1(t)$. Moreover, for any vector $\mathsf{s}(t)$ (and $\mathsf{i}(t)$, $\mathsf{o}(t)$), 
we use the notation $\mathsf{s}^i$ (and $\mathsf{i}^j$, $\mathsf{o}^k$) to present its value, where the superscript $i$ (and $j$, $k$) is equal to the binary number $\mathsf{s}_1 (t) \mathsf{s}_2 (t) \mathsf{s}_3 (t)$ (and $\mathsf{i}_1 (t)$, $\mathsf{o}_1 (t) \mathsf{o}_2 (t)$, respectively). Then the updating rules of it can be presented by its simplified form (Fig.~\ref{fig:7}). And in the rest of this paper, all the updating rules will be presented by their simplified  forms.


 %
\end{example}

\subsection{Control theoretical properties of \BCNs}
In this subsection, we introduce the notations controllability, observability and identifiability of {\BCNs} and their relations. 
In particular, we will give a summary about the existing work on observability in order to motivate our work. To this end, first define some notations below.



Given a {\BCN} $\BB = (I,S,O, E, \sigma, \rho)$, let $\mathcal{I}$, $\mathcal{S}$ and $\mathcal{O}$ be the sets of all possible inputs, states and outputs of $\BB$, respectively. Then a {\em timed run} of $\BB$ can be defined as a triple $R[t]=( \mathsf{I}[t], \mathsf{S}[t],  \mathsf{O}[t])$, where $ \mathsf{I}[t]\in \mathcal{I}^{t+1}$, $ \mathsf{S}[t]\in \mathcal{S}^{t+1}$ and $ \mathsf{O}[t]\in \mathcal{O}^{t+1}$  such that for $ \mathsf{I}[t]=  \mathsf{i}(0)\ldots  \mathsf{i}(t)$,  $ \mathsf{S}[t]=  \mathsf{s}(0)\ldots  \mathsf{s}(t)$ and $ \mathsf{O}[t]=  \mathsf{o}(0)\ldots  \mathsf{o}(t)$ 
\begin{itemize}
\item  $\forall t'=0,\ldots,t\cdot \mathsf{o}(t')=\rho(\mathsf{s}(t'))$, and
\item $\forall t'=1,\ldots,t\cdot \mathsf{s}(t')=\sigma( \mathsf{i}(t'-1),\mathsf{s}(t'-1))$.
\end{itemize}

		Secondly, we define the following two functions which define,  for any interval $[t_0,t+1]$ of observing time from $t_0$ to $t+1$,  the sequence of states and sequence of outputs produced, respectively,  in the interval by a state $\mathsf{s}(t_0)$ at time $t_0$ and a sequence of inputs in the interval $[t_0,t]$. 

\begin{equation}
\begin{split}
F^{[t_0,t]}&: \mathcal{S}\times \mathcal{I}^{(t-t_0 +1)} \mapsto \mathcal{S}^{(t-t_0 +2)}\\
F^{[t_0,t]}&(\mathsf{s}, \mathsf{i}(t_0)\ldots \mathsf{i}(t))= \mathsf{s}(t_0)\ldots \mathsf{s}(t+1)\\
 \end{split}
\end{equation}
 \begin{equation}
\begin{split}
 H^{[t_0,t]}&: \mathcal{S}\times \mathcal{I}^{(t-t_0 +1)} \mapsto \mathcal{O}^{(t-t_0 +2)}\\
H^{[t_0,t]}&(\mathsf{s}, \mathsf{i}(t_0)\ldots \mathsf{i}(t))= \mathsf{o}(t_0)\ldots \mathsf{o}(t+1)\\
\end{split}
\end{equation}

\[
\begin{array}{llllll}
\mbox{such that the following conditions are satisfied.}\\
 (\mathsf{s}(t_0)=\mathsf{s})\wedge \\
 \forall t'=(t_0+1),\ldots,(t+1)\cdot (\mathsf{s}(t')=\sigma(\mathsf{i}(t'-1),\mathsf{s}(t'-1)))\wedge \\
 \forall t'=t_0,\ldots,(t+1)\cdot (\mathsf{o}(t')=\rho(\mathsf{s}(t'))
\end{array}
\]

These two functions generalize the two functions given in ~\cite{Zhang2016Observability} for observability,  where only the special case of $F^{[0,t]}$ and $H^{[0,t]}$ are given. Their extensions will be used when we present the the new observability. 

Then we can define the following properties.

\begin{definition}[Controllability]
	A \BCN\ is controllable if for any two distinct states $\mathsf{s}$, $\mathsf{s}' \in \mathcal{S}$, there is an input sequence $\mathsf{I}[t]\in\mathcal{I}^{t+1}$ for some $t\in\mathbb{T}$, such that  $F^{[0,t]}(\mathsf{s}, \mathsf{I}[t])=\mathsf{s}(0) \ldots\, \mathsf{s}(t+1)$ and $\mathsf{s}(t+1)=\mathsf{s}'$ (\cite{cheng2009controllability}).
\end{definition}

Thus, if a \BCN\  $\BB$ is  controllable,  any state $\mathsf{s}'$ is reachable from any initial state $\mathsf{s}$, and  we use an input  sequence $\mathsf{I}[t]$ make $\BB$ reach $\mathsf{s}'$ from $\mathsf{s}$.

\begin{definition} [Observability]
We define the four types of observability below. 
\begin{enumerate}
  \item The {\sf Type-I} observability is that, a \BCN\ is observable if for every initial state $\mathsf{s}\in \mathcal{S}$, there exists an input sequence $\mathsf{I}[t]\in\mathcal{I}^{t+1}$ for some $t\in\mathbb{T}$, such that for all states $\mathsf{s}'$, $H^{[0,t]}(\mathsf{s}',\mathsf{I}[t])\neq H^{[0,t]}(\mathsf{s}, \mathsf{I}[t])$ if $\mathsf{s}\ne\mathsf{s}'$ (\cite{cheng2009controllability}).
  \item The  {\sf Type-II} observability is that, a \BCN\ is observable if for every two distinct initial states $\mathsf{s}$, $\mathsf{s}' \in \mathcal{S}$, there is an input sequence $\mathsf{I}[t]\in\mathcal{I}^{t+1}$ for some $t\in\mathbb{T}$, such that $H^{[0,t]}(\mathsf{s}',\mathsf{I}[t])\neq H^{[0,t]}(\mathsf{s}, \mathsf{I}[t])$ (\cite{Zhao2010Input}).
  \item The {\sf Type-III} observability is that, a \BCN\ is observable if there exists an input sequence $\mathsf{I}[t]\in\mathcal{I}^{t+1}$ for some $t\in\mathbb{T}$, such that for any two distinct states $\mathsf{s}$, $\mathsf{s}' \in \mathcal{S}$, $H^{[0,t]}(\mathsf{s}',\mathsf{I}[t])\neq H^{[0,t]}(\mathsf{s}, \mathsf{I}[t])$ (\cite{Cheng2011Identification}).
  \item The {\sf Type-IV} observability is that, a \BCN\ is observable, if there is a  natural number $N$, such that for every input sequence $\mathsf{I}[t]\in\mathcal{I}^{t+1}$ with $t\ge N$, $H^{[0,t]}(\mathsf{s}',\mathsf{I}[t])\neq H^{[0,t]}(\mathsf{s}, \mathsf{I}[t])$ holds for any two distinct states $\mathsf{s}$, $\mathsf{s}' \in \mathcal{S}$ (\cite{Fornasini2013Observability}).
\end{enumerate} 
\end{definition}

\begin{definition} [Reconstructibility]
We define the two types of observability below. 
\begin{enumerate}
\item The {\sf Type-I} reconstructibility is that, a \BCN\ is reconstructible, if there is a  natural number $N$, such that for every input sequence $\mathsf{I}[t]\in\mathcal{I}^{t+1}$ with $t\ge N$, $H^{[0,t]}(\mathsf{s}',\mathsf{I}[t])\neq H^{[0,t]}(\mathsf{s}, \mathsf{I}[t])$ holds for any two distinct states $\mathsf{s}$, $\mathsf{s}' \in \mathcal{S}$ if  their corresponding current states $\mathsf{s}(t+1)$ and $\mathsf{s}'(t+1)$ are different (\cite{Fornasini2013Observability}).
  \item The {\sf Type-II} reconstructibility is that, a \BCN\ is reconstructible if there exists an input sequence $\mathsf{I}[t]\in\mathcal{I}^{t+1}$ for some $t\in\mathbb{T}$, such that for any two distinct states $\mathsf{s}$, $\mathsf{s}' \in \mathcal{S}$, $H^{[0,t]}(\mathsf{s}',\mathsf{I}[t])\neq H^{[0,t]}(\mathsf{s}, \mathsf{I}[t])$  if their corresponding current states $\mathsf{s}(t+1)$ and $\mathsf{s}'(t+1)$ are different (\cite{Zhang2015A}).
  
\end{enumerate} 
\end{definition}

\begin{definition}[Identifiability]%
A \BCN\ is identifiable if  there exists an input sequence $\mathsf{I}[t]\in\mathcal{I}^{t+1}$ for some $t \in \mathbb{T}$ such that the updating rules
	\begin{equation*}
    		\begin{split}
		\mathsf{s}(t+1)=&\sigma(\mathsf{i}(t),\mathsf{s}(t))\\
		\mathsf{o}(t)=&\rho(\mathsf{s}(t))
		\end{split}
	\end{equation*}
	can be constructed by $\mathsf{I}[t]$ and its corresponding output sequence  $\mathsf{O}[t+1]$ (\cite{Cheng2011Identification}). 
\end{definition}


\subsubsection*{Remark} As mentioned in \cite{Cheng2011Identification}, to identify a \BCN, the numer of its state-nodes is required. Moreover, a \BCN\  $\BB$ with updating rules 
\begin{equation*}
    		\begin{split}
		\mathsf{s}(t+1)=&\sigma(\mathsf{i}(t),\mathsf{s}(t))\\
		\mathsf{o}(t)=&\rho(\mathsf{s}(t))
		\end{split}
	\end{equation*}
and a \BCN\  $\BB'$ with updating rules 
\begin{equation*}
\begin{split}
\mathsf{s}(t+1)=&\sigma'(\mathsf{i}(t),\mathsf{s}(t))=f(\sigma(\mathsf{i}(t),f^{-1}(\mathsf{s}(t))))\\
\mathsf{o}(t)=&\rho'(\mathsf{s}(t))=\rho(f^{-1}(\mathsf{s}(t)))
\end{split}
\end{equation*}  
are not distinguishable by any input-output data if the $f:\mathbb{B}^m\mapsto \mathbb{B}^m$ is a bijective function from the  $m$-dimension vectors to the  $m$-dimension vectors of Boolean values, and $f^{-1}$ is the inverse function of $f$. Thus, precisely speaking, what we identify is the equivalence class of updating rules: $$\{(f(\sigma(\mathsf{i}(t),f^{-1}(\mathsf{s}(t)))),\rho(f^{-1}(\mathsf{s}(t))))| f:\mathbb{B}^m \mapsto \mathbb{B}^m\},$$ not the specific updating rules $(\sigma, \rho)$. With the equivalence class, we can further identify the updating rules by the \BCN's structure.

In conclusion, the controllability and observability (or reconstructibility) are proposed for researching if it is possible and how to control or determine the initial state (or current state) of a \BCN, respectively, when we know its updating rules. While, the identifiability is proposed for researching if it is possible and how to find the equivalence class of updating rules by its inputs and outputs and size. These properties are also closely related. The {\sf Type-III} observability and {\sf Type-IV} observability imply the {\sf Type-II} reconstructibility and {\sf Type-I} reconstructibility, respectively. There is a theorem proposed in~\cite{Cheng2011Identification} which states that a \BCN\ is identifiable iff  it satisfies controllability and the {\sf Type-III} observability. While, in~\cite{inproceedings}, researchers state that controllability together with the {\sf Type-II} reconstructibility is the sufficient and necessary condition of identifiability.

%% file: sec/onlineobservability.tex
\section{The online observability of \BCNs}
\label{sec:online}
Different from the proposition presented in~\cite{Cheng2011Identification} which states that we can determine a \BCN's inital state by doing one experiment iff it satisfies the {\sf Type-III} observability.
In this section, we define the online observability to present the single-experiment observability for \BCNs, i.e. the property that the initial state of a \BCN\ can be determined without resetting. After that, we present the relationship between the online observability and existing four types of observability. 

To define the online observability, we start with the derivation function $\Ded( \mathsf{S},  \mathsf{i},  \mathsf{o})$. We write 

\begin{equation*}
\begin{split}
\xi :  (\mathcal{I}\cup\{\varepsilon\}) \times\mathcal{S} \mapsto\mathcal{S},\ \xi (\mathsf{i}, \mathsf{s})= \left\{
\begin{array}{rcl}
\sigma( \mathsf{i}, \mathsf{s})      &      & {\mathsf{i}\neq \varepsilon}\\
\mathsf{s}       &      & {\mathsf{i}= \varepsilon}
\end{array} \right. 
\end{split}
\end{equation*}
where $\varepsilon$ presents empty input. Then the derivation function $\Ded( \mathsf{S},  \mathsf{i},  \mathsf{o})$ can be defined as follows.

\begin{equation}
\begin{split}
\Ded&:2^{\mathcal{S}}\times(\mathcal{I}\cup\{\varepsilon\})\times(\mathcal{O}\cup\{\varepsilon\})\mapsto2^{\mathcal{S}}\\
\Ded&( \mathsf{S},  \mathsf{i},  \mathsf{o})= \left\{
\begin{array}{rcl}
\{  \xi(\mathsf{i}, \mathsf{s} ) \mid  \mathsf{s} \in \mathsf{S},  \rho(\xi(\mathsf{i}, \mathsf{s} ))=\mathsf{o}\}     &      & {\mathsf{o}\neq \varepsilon}\\
\{  \xi(\mathsf{i}, \mathsf{s} ) \mid  \mathsf{s} \in \mathsf{S}\}       &      & {\mathsf{o}= \varepsilon}
\end{array} \right. 
\end{split}
\end{equation}
where  $\varepsilon$ presents empty input or empty output.

 Secondly, 
we recursively define the following function to present how to derive the set $\mathsf{S}(t)$ of possible valuations of a \BCN's state $\mathsf{s}(t)$ at time $t$ by its input sequence $\mathsf{i}(0)\ldots \mathsf{i}(t-1)$ and output sequence $\mathsf{o}(0)\ldots \mathsf{o}(t)$.


 \begin{equation}
\begin{split}
 G^{[t]}&:\mathcal{I}^{t}\times  \mathcal{O}^{t+1} \mapsto   2^{\mathcal{S}}\\
G^{[t]}&( \mathsf{i}(0)\ldots \mathsf{i}(t-1), \mathsf{o}(0)\ldots \mathsf{o}(t))= \{\mathsf{s}^1,\ldots, \mathsf{s}^k\}\\
\end{split}
\end{equation}
such that the following conditions are satisfied. 
\begin{itemize}
 \item When $t= 0$, $\mathsf{i}(0)\ldots\mathsf{i}(t-1)=\varepsilon$ and \\ 
 $\{\mathsf{s}^1 ,\ldots, \mathsf{s}^k \}=\Ded(\mathcal{S},\varepsilon,\mathsf{o}(0))$;
 \item when $t> 0$, $\{\mathsf{s}^1,\ldots, \mathsf{s}^k\}=\Ded(G^{[t-1]}(\mathsf{i}(0)\ldots \mathsf{i}(t-2), \mathsf{o}(0)\ldots \mathsf{o}(t-1)),\mathsf{i}(t-1),\mathsf{o}(t))$. 
 \end{itemize}
 
That is the set $\mathsf{S}(t)=G^{[t]}( \mathsf{i}(0)\ldots \mathsf{i}(t-1), \mathsf{o}(0)\ldots \mathsf{o}(t))$, and for any $t_0 :0<{t_0}\le t$, $\mathsf{S}({t_0})=\Ded(\mathsf{S}({t_0}-1),\mathsf{i}({t_0}-1),\mathsf{o}({t_0}))$.

\begin{example}
In the \BCN\ in {\em Example \ref{exa:2}}, if $t=2$, $\mathsf{i}(0)\ldots\mathsf{i}(t-1)=\mathsf{i}^1 \mathsf{i}^1$ and $\mathsf{o}(0)\ldots\mathsf{o}(t)=\mathsf{o}^1 \mathsf{o}^2 \mathsf{o}^3$, then 
\begin{equation*}
\begin{split}
&\mathsf{S}(t-2)=G^{[0]}(\varepsilon,\mathsf{o}^1)=\Ded\left(\mathcal{S},\varepsilon,\mathsf{o}^1\right)=\{\mathsf{s}^1,\mathsf{s}^2,\mathsf{s}^3\},\\
&\mathsf{S}(t-1)=G^{[1]}(\mathsf{i}^1,\mathsf{o}^1 \mathsf{o}^2)=\Ded\left(\mathsf{S}(t-2),\mathsf{i}^1,\mathsf{o}^2\right)=\{\mathsf{s}^4,\mathsf{s}^5\},\\
&\mathsf{S}(t)=G^{[2]}(\mathsf{i}^1 \mathsf{i}^1,\mathsf{o}^1 \mathsf{o}^2 \mathsf{o}^3)=\Ded\left(\mathsf{S}(t-1),\mathsf{i}^1,\mathsf{o}^3\right)=\{\mathsf{s}^6\}.
 \end{split}
\end{equation*}
 \label{exa:8}
 \end{example}

Then, we can get a conclusion that the state $\mathsf{s}(t)$ can be determined in $k$ steps without resetting iff 
\begin{itemize}
 \item  $|\mathsf{S}(t+k)|=1$, i.e. the $\mathsf{s}(t+k)$ is determined, and 
 \item   for every $t_0: t+1\le t_0 \le t+k$, for every $\mathsf{s}(t_0)\in\mathsf{S}(t_0)$ there exists only one $\mathsf{s}'(t_0-1)\in\mathsf{S}(t_0-1)$ satisfies that $\mathsf{s}(t_0)=\sigma(\mathsf{i}(t_0-1), \mathsf{s}'(t_0-1))$.
 \end{itemize}

Thus, as the next step, we define a function $\Ks(\mathsf{S})$ to depict the number of steps we need to determine $\mathsf{s}(t)$ if $\mathsf{S}(t)=\mathsf{S}$.
\begin{equation}
\begin{split}
\Ks:(2^{\mathcal{S}}-\emptyset) \mapsto (\mathbb{T}\cup\{\infty\})
 \end{split}
\end{equation}
which satisfies that

if $|\mathsf{S}|=1$, then $\Ks(\mathsf{S})=0$;

if $|\mathsf{S}|>1$
 \begin{itemize}
 \item  if there is an input $\mathsf{i} \in \mathcal{I}$ such that 
 \begin{itemize}
 \item  $|\Ded(\mathsf{S},\mathsf{i},\varepsilon)|=|\mathsf{S}|$, and 
 \item   $\forall\mathsf{o} \in \mathcal{O}\cdot\Ded(\mathsf{S},\mathsf{i},\mathsf{o})\ne \emptyset\to\Ks(\Ded(\mathsf{S},\mathsf{i},\mathsf{o}))\ne \infty$,
 \end{itemize} 
 then 
 \begin{equation*}
\begin{split}
\Ks(\mathsf{S})=&1+\\
&\min\limits_{\mathsf{i}' \in\{ \mathsf{i}| |\Ded(\mathsf{S},\mathsf{i},\varepsilon)|=|\mathsf{S}|\}}\max\limits_{\mathsf{o}'\in\{ \mathsf{o}| \Ded(\mathsf{S},\mathsf{i},\mathsf{o})\ne \emptyset\}}\Ks(\Ded(\mathsf{S},\mathsf{i}',\mathsf{o}'))
 \end{split}
\end{equation*}
 \item  otherwise, $\Ks(\mathsf{S})=\infty$. 
 
 \end{itemize}

\begin{example}\label{exa:9}
In the \BCN\ in {\em Example \ref{exa:2}}, if $\mathsf{o}(0)=\mathsf{o}^2$ then \[\mathsf{S}(0)=\Ded\left(\mathcal{S},\varepsilon,\mathsf{o}^2\right)=\{\mathsf{s}^4,\mathsf{s}^5\}.\] 
As $|\mathsf{S}(0)|=2>1$, and there exists an input $\mathsf{i}^1$ such that 
 \begin{itemize}
 \item  $|\Ded\left(\mathsf{S}(0),\mathsf{i}^1,\varepsilon\right)|=|\{\mathsf{s}^{2},\mathsf{s}^6\}|=|\mathsf{S}(0)|$,
 \item   and for each $\mathsf{o}\in \mathcal{O}$ that $\Ded(\mathsf{S}(0),\mathsf{i},\mathsf{o})\ne \emptyset$
  \begin{itemize}
  \item   $\Ks(\Ded\left(\mathsf{S}(0),\mathsf{i}^1,\mathsf{o}^1\right))=\Ks(\{\mathsf{s}^2\})=0$;
 \item  $\Ks(\Ded\left(\mathsf{S}(0),\mathsf{i}^1,\mathsf{o}^3\right))=\Ks(\{\mathsf{s}^{6}\})=0$,
 \end{itemize}
 \end{itemize}
we have  
\begin{equation*}
\begin{split}
\Ks&(\mathsf{S}(0))=1+\\
&\min\limits_{\mathsf{i}' \in\{ \mathsf{i}| |\Ded(\mathsf{S}(0),\mathsf{i},\varepsilon)|=|\mathsf{S}(0)|\}}\max\limits_{\mathsf{o}'\in\{ \mathsf{o}| \Ded(\mathsf{S}(0),\mathsf{i},\mathsf{o})\ne \emptyset\}}\Ks(\Ded(\mathsf{S}(0),\mathsf{i}',\mathsf{o}'))\\
&=1+0=1
 \end{split}
\end{equation*}
i.e. the $\mathsf{s}(0)$ can be determined at time $1$ without resetting. 
\end{example}  




Then, 
the online observability can be defined as follows.
\begin{definition}[Online Observability of  BCNs]
 A \BCN\ is online observable
if for every $\mathsf{o}\in  \mathcal{O}$, $\Ded(\mathcal{S},\varepsilon, \mathsf{o})\ne\emptyset$ implies $\Ks(\Ded(\mathcal{S},\varepsilon,\mathsf{o}))\ne \infty$.
\end{definition}

That is the initial state $\mathsf{s}(0)$ of a \BCN\ can be determined without resetting iff for every possible $\mathsf{S}(0)$, $\Ks(\mathsf{S}(0))\ne \infty$. 


In order to make this definition more convincing, we illustrate how to determine the initial state of a \BCN\ with online observability below. To this end, we define $\Ri(\mathsf{S})$ to depict the set of inputs we can choose from at time $t$ in the process of determining the initial state, if $\mathsf{S}(t)=\mathsf{S}$.

\begin{equation}
\begin{split}
\Ri&:(2^{\mathcal{S}}-\emptyset) \mapsto2^{\mathcal{I}}\\
\Ri&(\mathsf{S})=\{\mathsf{i}\in \mathcal{I}~|~|\Ded\left(\mathsf{S},\mathsf{i},\varepsilon\right)|=|\mathsf{S}|, \\
&~~~~~~~~\forall \mathsf{o} \in \mathcal{O}\cdot \Ded(\mathsf{S},\mathsf{i},\mathsf{o})\ne \emptyset \rightarrow \Ks(\Ded(\mathsf{S},\mathsf{i},\mathsf{o}))\ne \infty\}
\end{split}
\end{equation}

Then we can determine the initial state of a \BCN\ $\BB$ with online observability by followng procedures.

\begin{description}
	\item[Step 1]  Derive the set $\mathsf{S}(0)$ of possible valuations of initial state $\mathsf{s}(0)$ by the output $\mathsf{o}(0)$ of $\BB$, i.e. $\mathsf{S}(0)=\Ded\left(\mathcal{S},\varepsilon,\mathsf{o}(0)\right)$, and set the set variable $\mathsf{S}=\mathsf{S}(0)$ by $\mathsf{S}(0)$.
	\item[Step 2] Input to the \BCN\ $\BB$ with an input $\mathsf{i}\in\Ri(\mathsf{S})$, and run $\BB$ to generate the new output $\mathsf{o}(t)$. 
	\item[Step 3] Determine the new $\mathsf{S}(t)$ by the input $\mathsf{i}$, output $\mathsf{o}(t)$ and $\mathsf{S}$, i.e. $\mathsf{S}(t)=\Ded(\mathsf{S},\mathsf{i},\mathsf{o}(t))$, and update the set $\mathsf{S}=\mathsf{S}(t)$ by $\mathsf{S}(t)$.
	\item[Step 4] If the cardinal number $|\mathsf{S}|=1$, then return $\mathsf{s}(0)$ which is determined by the output sequence $\mathsf{O}[t]=\mathsf{o}(0)\ldots\mathsf{o}(t)$ and input sequence $\mathsf{I}[t-1]=\mathsf{i}(0)\ldots\mathsf{i}(t-1)$ as the initial state. 
	 Otherwise, go to Step 2.
\end{description}
  In Step 4, when $|\mathsf{S}|=1$, for the input sequence $\mathsf{I}[t-1]$, there is only one state $\mathsf{s}$ whose corresponding output sequence $H^{[0,t-1]}(\mathsf{s},\mathsf{I}[t-1])$ equals to the output sequence $\mathsf{O}[t]$ of $\BB$, thus, we can determine the initial state by $\mathsf{I}[t]$ and $\mathsf{O}[t]$, i.e. $\mathsf{s}(0)=\mathsf{s}$. 

\begin{example}\label{exa:3} The \BCN\ in {\em Example \ref{exa:2}} is a convincing example. In this \BCN, 
we have $\Ks(\Ded\left(\mathcal{S},\varepsilon, \mathsf{o}^0\right))=0$; $\Ks(\Ded\left(\mathcal{S},\varepsilon, \mathsf{o}^1\right))=2$; $\Ks(\Ded\left(\mathcal{S},\varepsilon, \mathsf{o}^2\right))=1$; $\Ks(\Ded\left(\mathcal{S},\varepsilon, \mathsf{o}^3\right))=1$,  
Thus, it is online observable, and we can determine its initial state by above procedures without resetting.

Moreover, this \BCN\ does not satisfy the {\sf Type-III} observability because for any $t\in \mathbb{T}$, 
\begin{itemize}
\item for any input sequence $\mathsf{I}[t]$ starting with $\mathsf{i}^0$, \\$H^{[0,t]}(\mathsf{s}^1,\mathsf{I}[t])= H^{[0,t]}(\mathsf{s}^{2}, \mathsf{I}[t])$;
  \item for any input sequence $\mathsf{I}[t]$ starting with $\mathsf{i}^1$, \\$H^{[0,t]}(\mathsf{s}^6,\mathsf{I}[t])= H^{[0,t]}(\mathsf{s}^7, \mathsf{I}[t])$.
\end{itemize} 
i.e. for any $t\in \mathbb{T}$, there is not any input sequence $\mathsf{I}[t]\in\mathcal{I}^{[t]}$ which satisfies that for any two distinct states $\mathsf{s}$, $\mathsf{s}' \in \mathcal{S}$, $H^{[0,t]}(\mathsf{s}',\mathsf{I}[t])\neq H^{[0,t]}(\mathsf{s}, \mathsf{I}[t])$. 

\end{example}

We call the observability we propose online observability because in above procedure, we choose the input $\mathsf{i}(t)$ at evey time $t$ based on the information of the inputs and outputs of \BCN\ we have collected so far, i.e. $\mathsf{i}\in\Ri(\mathsf{S}(t))$. In contrast, we call the {\sf type-III} observability {\em offline observability}, because in the algorithm of {\sf type-III} observability, we determine the initial state $\mathsf{s}(0)$ of a \BCN\ by its recorded output sequence $\mathsf{O}[t]= H^{[0,t-1]}(\mathsf{s}(0),\mathsf{I}[t-1])=\mathsf{o}(0)\ldots\mathsf{o}(t)$ after we input $\mathsf{I}[t-1]=\mathsf{i}(0)\ldots\mathsf{i}(t-1)$. That is we do not interfere with \BCN\ except for the logging of its inputs and outputs.

\begin{figure}
		\begin{center}
			\includegraphics[width=5cm]{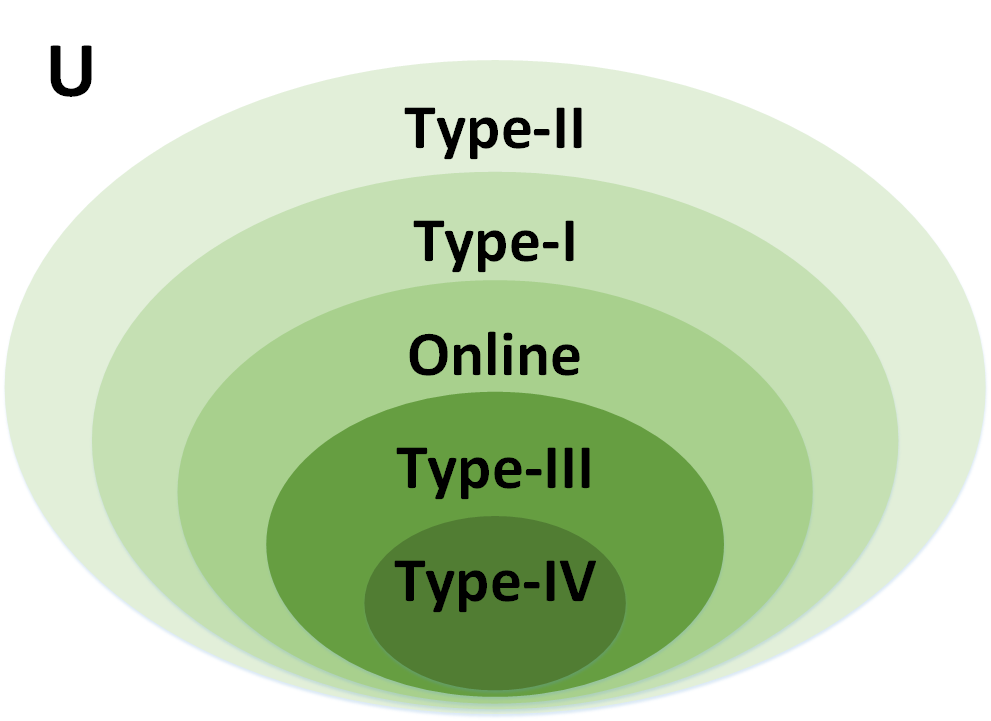}    
			\caption{The relationship between five types of observability.} 
			\label{fig:11}
		\end{center}
	\end{figure}

\begin{lemma}
The {\sf Type-III} observability implies the online observability.
\label{lemm:4}
\end{lemma}

\begin{lemma}
The online observability implies the  {\sf Type-I} observability.
\label{lemm:3}
\end{lemma}

Moreover, we propose two lemmas for the implication relationship between the {\sf Type-I \& III} observability and online observability. The proofs of all lemmas are shown in the extended version (\cite{WuOnline}) of this article due to space limits. Moreover, it is noted in ~\cite{Zhang2016Observability} that the {\sf Type-I} observability is stronger than the {\sf Type-II} observability; the {\sf Type-III} observability is stronger than the {\sf Type-I} observability;  and the {\sf Type-IV} observability is the strongest. Therefore, the implication relationship of five type of observability can be shown by Fig.~\ref{fig:11}, in which, the area which is labelled with a type of observability presents the set of \BCNs\ which satisfy this type of observability, and  the area which is labelled with ``U" presents the set of all of the \BCNs.




%% file: sec/determining.tex

\section{Algorithms for online observability}
\label{sec:deter}
In this section, we propose decision algorithm for the online observability. To decide whether a \BCN\ $\BB$ satisfies online observability, one needs to determine $\Ks(\Ded(\mathcal{S},\varepsilon,\mathsf{o}))$ of every non-empty $\Ded(\mathcal{S},\varepsilon, \mathsf{o})$. 

We begin with the input-labelled graph $\mathcal{G}=(\mathcal{V}, \mathcal{E}, \mathcal{L})$.

\begin{definition}[Input-labelled Graph]
Let $\mathcal{V}$, $\mathcal{E}$ and $\mathcal{L}$ be the vertex set, the edge set and the labelling function of an input-labelled graph $\mathcal{G}=(\mathcal{V}, \mathcal{E}, \mathcal{L})$. $\mathcal{G}$ is called the input-labelled graph of the \BCN\, if 
\begin{itemize}
\item  $\mathcal{V}=\{\mathsf{S}\in(\bigcup\limits_{\mathsf{o}\in\mathcal{O}} 2^{\Ded(\mathcal{S},\varepsilon,\mathsf{o} )}-\emptyset)~|~\ \Ks(\mathsf{S})\ne \infty\}$;
\item  $\mathcal{E}=\{(\mathsf{S}_1,\mathsf{S}_2)\in \mathcal{V}\times \mathcal{V}~|~\mathsf{S}_2\in\{\Ded(\mathsf{S}_1,\mathsf{i},\mathsf{o})~|~\mathsf{i}\in\Ri(\mathsf{S}_1),\mathsf{o}\in  \mathcal{O}\}\}$;
\item  $\mathcal{L}:\mathcal{E}\mapsto 2^{\mathcal{I}}$, $\mathcal{L}(\mathsf{S}_1,\mathsf{S}_2)=\{\mathsf{i}\in \Ri(\mathsf{S}_1)~|~\mathsf{S}_2\in\{\Ded(\mathsf{S}_1,\mathsf{i},\mathsf{o})~|~\mathsf{o}\in  \mathcal{O}\}\}$.
 \end{itemize}
\end{definition}

\begin{example}
The input-labelled graph of the \BCN\ in {\em Example \ref{exa:2}} is shown in Fig.~\ref{fig:4}. 
	\begin{figure}
		\begin{center}
			\includegraphics[width=5cm]{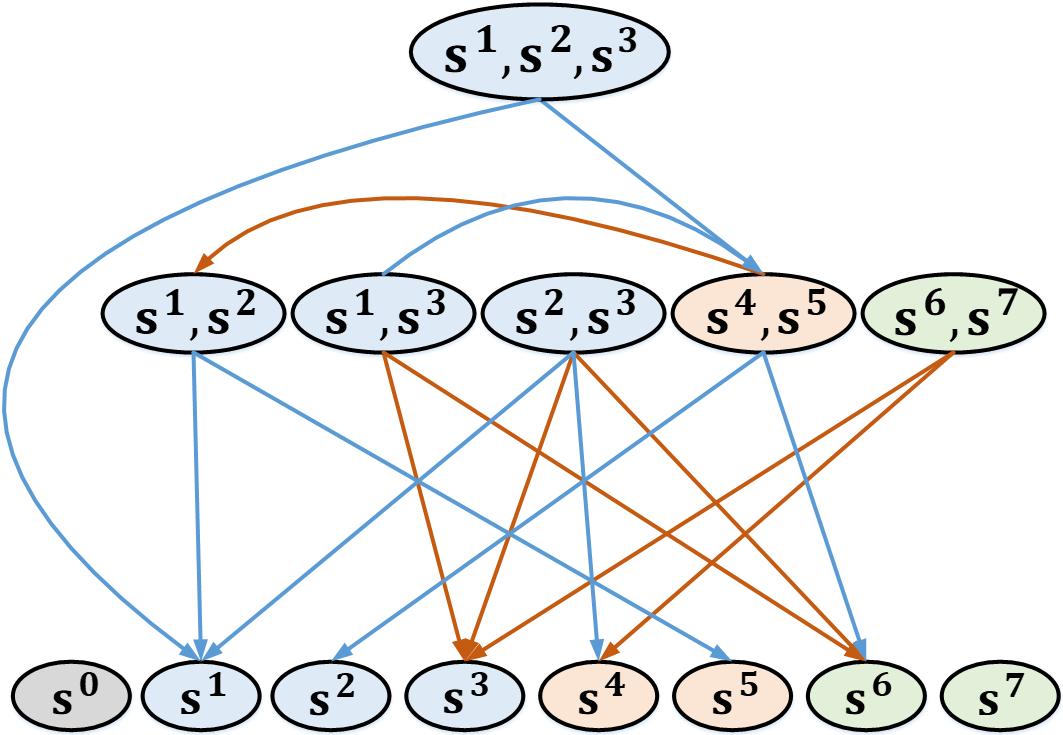}    
			\caption{The input-labelled graph, where the orange and blue edges are labelled with $\{\mathsf{i}^0\}$ and $\{\mathsf{i}^1\}$, respectively.} 
			\label{fig:4}
		\end{center}
	\end{figure}
 \label{exa:104}
\end{example}

Intuitively, in the input-labelled graph $\mathcal{G}=(\mathcal{V}, \mathcal{E}, \mathcal{L})$, $\mathcal{V}$ represents a set of the state sets that for every $\mathsf{S}\in \mathcal{V}$, $\Ks(\mathsf{S})\ne \infty$, and for every two distinct states $\mathsf{s},\mathsf{s}'\in\mathsf{S}$, they produce the same output; $\mathcal{E}$ represents the relationship between the state sets which belong to $\mathcal{V}$; and $\mathcal{L}$ labels every edge $\mathsf{e} \in\mathcal{E}$ with a set of inputs. 
Thus, we have a \BCN\ $\BB$ is online observable iff every non-empty $\Ded(\mathcal{S},\varepsilon, \mathsf{o})\in \mathcal{V}$. 



Secondly, we propose two lemmas for the funtions $\Ks(\mathsf{S})$ and $\Ri(\mathsf{S})$, respectively.

\begin{lemma}
For any two non-empty state sets $\mathsf{S}^{1}$ and $\mathsf{S}^{2}$, if $\mathsf{S}^{1}\subseteq\mathsf{S}^{2}$ and $\Ks(\mathsf{S}^{2})\ne\infty$, then $\Ks(\mathsf{S}^{1})\ne\infty$.
  \label{lemm:1}
\end{lemma}

\begin{lemma}
For any two non-empty state sets  $\mathsf{S}^{1}$ and $\mathsf{S}^{2}$, if $\mathsf{S}^{1}\subseteq\mathsf{S}^{2}$ and $\Ks(\mathsf{S}^{2})\ne\infty$, then $\Ri(\mathsf{S}^2)\subseteq\Ri(\mathsf{S}^1)$.
\label{lemm:2}
\end{lemma}

With {\em Lemma \ref{lemm:1}, \ref{lemm:2}} and input-labelled graph, we propose the algorithm (which is also shown in~\cite{WuOnline}) to determine the online observability. That is we construct the input-labelled graph $\mathcal{G}=(\mathcal{V}, \mathcal{E}, \mathcal{L})$ for a \BCN\ $\BB$ at first, and then check whether every non-empty $\Ded(\mathcal{S},\varepsilon, \mathsf{o})\in \mathcal{V}$.
 In the process of constructing the input-labelled graph, we construct the vertexes which consist of smaller numbers of states before constructing the vertexes which consist of greater numbers of states for the following two reasons.

\begin{itemize}
\item  If we find a set of states $\mathsf{S}\notin \mathcal{V}$, then as there exists an output $\mathsf{o}\in \mathcal{O}$ which satisfies that $\mathsf{S}\subseteq \Ded(\mathcal{S},\varepsilon, \mathsf{o})$, we have $\Ks(\Ded(\mathcal{S},\varepsilon, \mathsf{o}))= \infty$ and this \BCN\ $\BB$ is not online observable 
 based on the {\em Lemma \ref{lemm:1}}.
\item  And based on the {\em Lemma \ref{lemm:2}}, if we have determined $\Ri(\mathsf{S}')$ for every $\mathsf{S}'\subset\mathsf{S}$, we can determine the approximate scope of $\Ri(\mathsf{S})$. With the scope of $\Ri(\mathsf{S})$, we can determine the $\Ks(\mathsf{S})$ more easily.
 \end{itemize} 

%% file: sec/identification.tex
\section{Observability and identifiability}
\label{sec:iden}

In this section, we prove that a  \BCN\ is identifiable iff it has controllability and online observability.  

\begin{figure}
		\begin{center}
			\includegraphics[width=7cm]{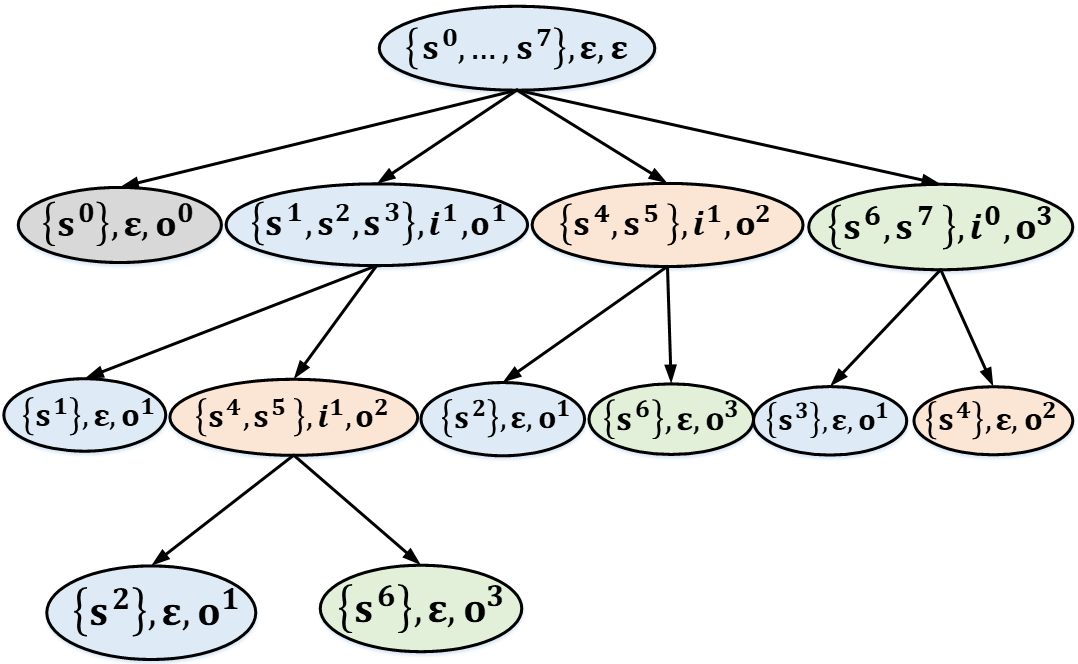}    
			\caption{The determining tree.} 
			\label{fig:3}
		\end{center}
	\end{figure}

We define the {\em determining tree} to illustrate the processes of determining the initial state at first. 
\begin{definition}[Determining tree]
If a \BCN\ $\BB$ satisfies the online observability then we can great at least one determining tree for it. In the tree, every node $\mathsf{n}$ is a variable which can take a set of states, an input and an output (the input and output can be $\varepsilon$), i.e. $\mathsf{n}=(\mathsf{S},\mathsf{i},\mathsf{o})$. If a node $\mathsf{n}$ is the root node then $\mathsf{n}=(\mathcal{S},\varepsilon,\varepsilon)$; if a node $\mathsf{n}=(\mathsf{S},\mathsf{i},\mathsf{o})$ is a leaf node then $|\mathsf{S}|=1$, $\mathsf{i}=\varepsilon$; if a node $\mathsf{n}=(\mathsf{S},\mathsf{i},\mathsf{o})$ is not a leaf node then its successor nodes form a set of nodes $\{\mathsf{n}_{[1]}=(\mathsf{S}_{[1]},\mathsf{i}_{[1]},\mathsf{o}_{[1]}),\ldots,\mathsf{n}_{[k]}=(\mathsf{S}_{[k]},\mathsf{i}_{[k]},\mathsf{o}_{[k]})\}$ that 
  \begin{itemize}
   \item  for every $\mathsf{n}_{[x]}=(\mathsf{S}_{[x]},\mathsf{i}_{[x]},\mathsf{o}_{[x]})\in\{\mathsf{n}_{[1]}=(\mathsf{S}_{[1]},\mathsf{i}_{[1]},\mathsf{o}_{[1]}),$\\$\ldots,\mathsf{n}_{[k]}=(\mathsf{S}_{[k]},\mathsf{i}_{[k]},\mathsf{o}_{[k]})\}$, $\mathsf{S}_{[x]}=\Ded(\mathsf{S},\mathsf{i},\mathsf{o}_{[x]})$, and $\mathsf{i}_{[x]}\in\Ri(\mathsf{S}_{[x]})$ if $\mathsf{n}_{[x]}$ is not a leaf node; and
   \item for any two distinct $\mathsf{n}_{[x]}=(\mathsf{S}_{[x]},\mathsf{i}_{[x]},\mathsf{o}_{[x]})$ and $\mathsf{n}_{[y]}=(\mathsf{S}_{[y]},\mathsf{i}_{[y]},\mathsf{o}_{[y]})\in\{\mathsf{n}_{[1]}=(\mathsf{S}_{[1]},\mathsf{i}_{[1]},\mathsf{o}_{[1]}),\ldots,\mathsf{n}_{[k]}=(\mathsf{S}_{[k]},\mathsf{i}_{[k]},\mathsf{o}_{[k]})\}$, $\mathsf{o}_{[x]}\ne\mathsf{o}_{[y]}$; and 
   \item $|\mathsf{S}_{[1]}|+,\ldots,|\mathsf{S}_{[k]}|=|\mathsf{S}|$.
 \end{itemize}
 Then we have the number of leaf nodes is equal to the number $2^m$ of all states of the \BCN\ $\BB$, where $m$ is the number of state-nodes.
\end{definition}
Intuitively, a path of the determining tree presents a possible determining process of the initial state when we choose a specific input $\mathsf{i}(t)$ from $\Ri(\mathsf{S}(t))$ for every $\mathsf{S}(t)$.
\begin{example}
As the \BCN\ in {\em Example \ref{exa:2}} satisfies the online observability, we can construct a determining tree (Fig.~\ref{fig:3}) for it. That we choose the specific inputs $\mathsf{i}^1$, $\mathsf{i}^0$, and $\mathsf{i}^1$ for the state sets  $\{\mathsf{s}^1,\mathsf{s}^2,\mathsf{s}^3\}$, $\{\mathsf{s}^6,\mathsf{s}^7\}$ and  $\{\mathsf{s}^4,\mathsf{s}^5\}$, respectively.
 \label{exa:101}
\end{example}  

Secondly, we define the {\em none state determining tree}.

\begin{figure}
		\begin{center}
			\includegraphics[width=5.5cm]{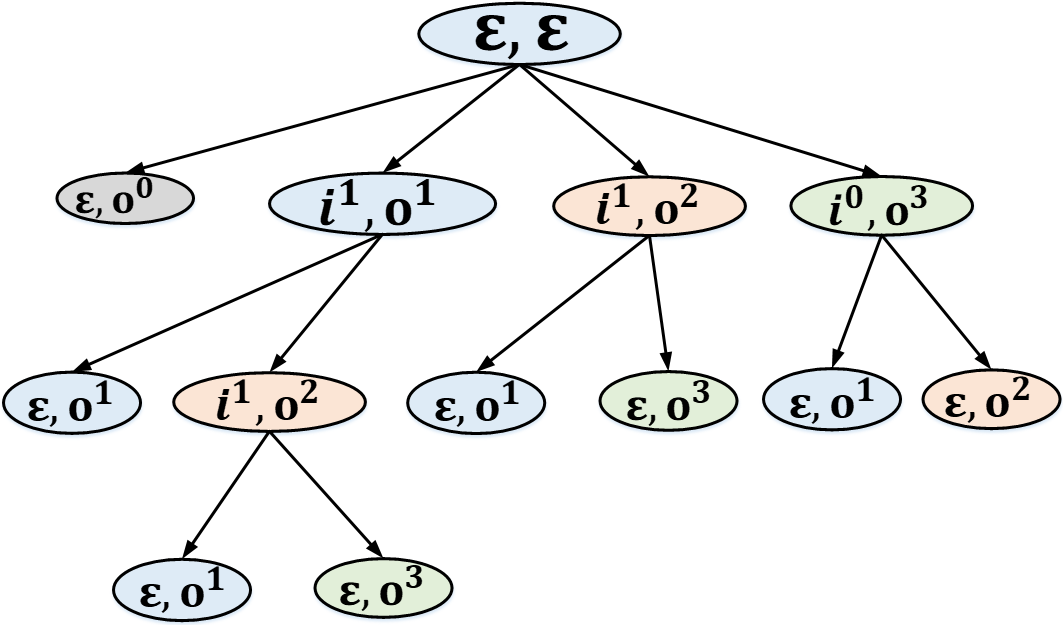}    
			\caption{The none state determining tree.} 
			\label{fig:3n}
		\end{center}
	\end{figure}

\begin{definition}[ None state determining tree]
 In the none state  determining tree, every node $\mathsf{n}$ is a variable which can take an input and an output (the input and output can be $\varepsilon$), i.e. $\mathsf{n}=(\mathsf{i},\mathsf{o})$. If a node $\mathsf{n}$ is the root node then $\mathsf{n}=(\varepsilon,\varepsilon)$; if a node $\mathsf{n}=(\mathsf{i},\mathsf{o})$ is a leaf node then $\mathsf{i}=\varepsilon$, and the number of leaf nodes is equal to $2^m$; if a node $\mathsf{n}=(\mathsf{i},\mathsf{o})$ is not a leaf node then its successor nodes form a set of nodes $\{\mathsf{n}_{[1]}=(\mathsf{i}_{[1]},\mathsf{o}_{[1]}),\ldots,\mathsf{n}_{[k]}=(\mathsf{i}_{[k]},\mathsf{o}_{[k]})\}$ that for any two distinct $\mathsf{n}_{[x]}=(\mathsf{i}_{[x]},\mathsf{o}_{[x]})$ and $\mathsf{n}_{[y]}=(\mathsf{i}_{[y]},\mathsf{o}_{[y]})\in\{\mathsf{n}_{[1]}=(\mathsf{i}_{[1]},\mathsf{o}_{[1]}),\ldots,\mathsf{n}_{[k]}=(\mathsf{i}_{[k]},\mathsf{o}_{[k]})\}$, $\mathsf{o}_{[x]}\ne\mathsf{o}_{[y]}$. 
\end{definition}
\begin{example}
 We construct a none state determining tree (Fig.~\ref{fig:3n}) by removing the set of states in Fig.~\ref{fig:3}. 
 \label{exa:102}
\end{example} 
With the determining tree and none state determining tree, we propose the following lemma.

\begin{figure}
		\begin{center}
			\includegraphics[width=6cm]{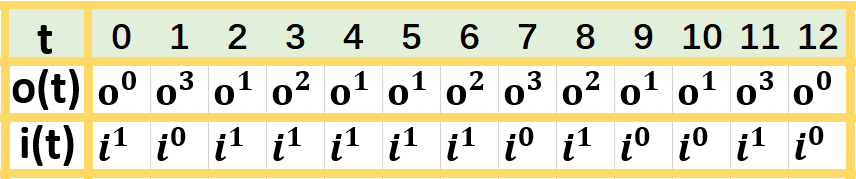}    
			\caption{The input-output data.} 
			\label{fig:6}
		\end{center}
	\end{figure}

\begin{figure}
		\begin{center}
			\includegraphics[width=6cm]{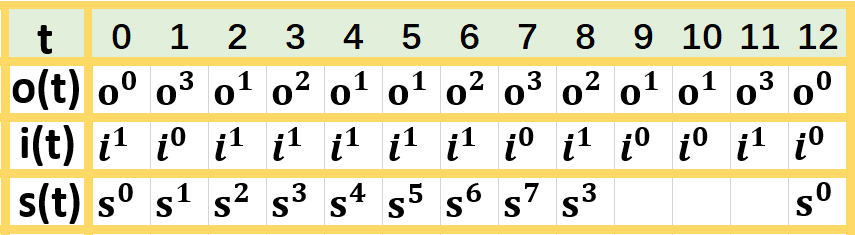}    
			\caption{The input-output-state data.} 
			\label{fig:6s}
		\end{center}
	\end{figure}

\begin{figure}
		\begin{center}
			\includegraphics[width=7cm]{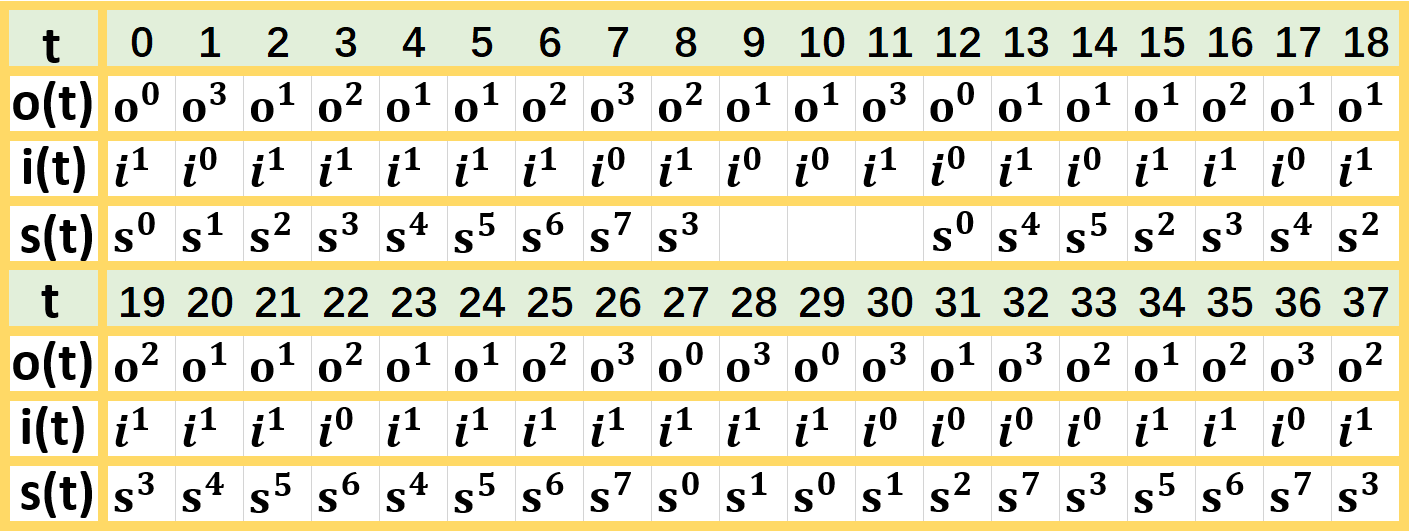}    
			\caption{The further constructed input-output-state data.} 
			\label{fig:6se}
		\end{center}
	\end{figure}

\begin{figure}
		\begin{center}
			\includegraphics[width=5.5cm]{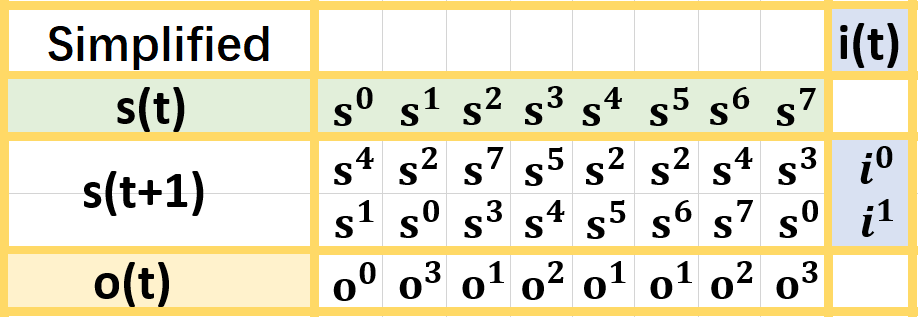}    
			\caption{The the updating rules of $\BB'$.} 
			\label{fig:7n}
		\end{center}
	\end{figure}

\begin{figure}
		\begin{center}
			\includegraphics[width=5cm]{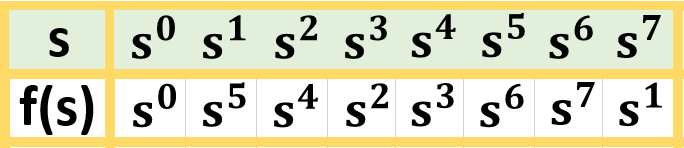}    
			\caption{The coordinate bijective function $f$.} 
			\label{fig:7t}
		\end{center}
	\end{figure}

\begin{lemma}
A \BCN\ is identifiable iff it satisfies the controllability and the online observability.
\label{lemm:5}
\end{lemma}
The proof of this lemma is shown in the extended version (\cite{WuOnline}) of this article due space limits. Shortly speaking, if a \BCN\ $\BB$ is online observable and controllable, we can construct a none state determining tree from its input-output data. With this tree, we can identify the updating rules $(\sigma',\rho')$ for a \BCN\ $\BB'$ which is equivalent to $\BB$. Thus this \BCN\ $\BB$ is identifiable. In order to make this lemma more convincing, we present the following example.
\begin{example}
As the \BCN\ $\BB$ in {\em Example \ref{exa:2}} satisfies the online observability and controllability, we can construct the input-output data (Fig.~\ref{fig:6}) from it.  

Firstly, we can construct a none state determining tree (Fig.~\ref{fig:3n}) by the $2^3$ pairs of input sequences and output sequences $(\varepsilon,\mathsf{o}^0)$, $(\mathsf{i}^0,\mathsf{o}^3 \mathsf{o}^1)$, $(\mathsf{i}^1\mathsf{i}^1,\mathsf{o}^1  \mathsf{o}^2\mathsf{o}^1)$, $(\mathsf{i}^1,\mathsf{o}^2\mathsf{o}^1)$, $(\mathsf{i}^1,\mathsf{o}^1 \mathsf{o}^1)$, $(\mathsf{i}^1\mathsf{i}^1,\mathsf{o}^1  \mathsf{o}^2\mathsf{o}^3)$, $(\mathsf{i}^1,\mathsf{o}^2\mathsf{o}^3)$, $(\mathsf{i}^0,\mathsf{o}^3 \mathsf{o}^2)$ which can be found in the input-output data. 
As these pairs can construct a none state determining tree, we have the $2^3$ sets of states $\{\mathsf{s}|\rho(\mathsf{s})=\mathsf{o}^0\}$, $\{\mathsf{s}|H^{[0,0]}(\mathsf{s},\mathsf{i}^0)=\mathsf{o}^3 \mathsf{o}^1\}$, $\{\mathsf{s}|H^{[0,1]}(\mathsf{s},\mathsf{i}^1\mathsf{i}^1)=\mathsf{o}^1  \mathsf{o}^2\mathsf{o}^1\}$, $\{\mathsf{s}|H^{[0,0]}(\mathsf{s},\mathsf{i}^1)=\mathsf{o}^2 \mathsf{o}^1\}$, $\{\mathsf{s}|H^{[0,0]}(\mathsf{s},\mathsf{i}^1)=\mathsf{o}^1 \mathsf{o}^1\}$, $\{\mathsf{s}|H^{[0,1]}(\mathsf{s},\mathsf{i}^1\mathsf{i}^1)=\mathsf{o}^1  \mathsf{o}^2\mathsf{o}^3\}$, $\{\mathsf{s}|H^{[0,0]}(\mathsf{s},\mathsf{i}^1)=\mathsf{o}^2 \mathsf{o}^3\}$ and $\{\mathsf{s}|H^{[0,0]}(\mathsf{s},\mathsf{i}^0)=\mathsf{o}^3 \mathsf{o}^2\}$ are disjoint. We regard them as the sets of states $\{\mathsf{s}^0\}$, $\{\mathsf{s}^1\}$, $\{\mathsf{s}^2\}$, $\{\mathsf{s}^3\}$, $\{\mathsf{s}^4\}$, $\{\mathsf{s}^5\}$, $\{\mathsf{s}^6\}$ and $\{\mathsf{s}^7\}$ of a \BCN\ $\BB'$ (which is equivalent to $\BB$), respectively (Fig.~\ref{fig:6s}). 
Then, we can construct the $\rho'$ (shown in Fig.~\ref{fig:7n}) for $\BB'$ from this input-output-state data.

Secondly, for every $\mathsf{s}$, for every $\mathsf{i}$, the $\mathsf{s}'=\sigma' (\mathsf{i},\mathsf{s})$ can be determined by the tree (Fig.~\ref{fig:3n}).
Moreover, from the input-output-state data (Fig.~\ref{fig:6s}), we have for any two distinct states $\mathsf{s}$ and $\mathsf{s}'$ of $\BB'$, there exists an input sequence which can make $\BB'$ reaches $\mathsf{s}$ from $\mathsf{s}'$. 
Therefore, we can further construct the input-output-state data (Fig.~\ref{fig:6se}), and construct the $\sigma'$ (Fig.~\ref{fig:7n}) for $\BB'$.

The \BCN\ $\BB'$ is equivalent to $\BB$ because the $(\sigma',\rho')$ satisfies  
\begin{equation*}
\begin{split}
\mathsf{s}(t+1)=&\sigma'(\mathsf{i}(t),\mathsf{s}(t))=f(\sigma(\mathsf{i}(t),f^{-1}(\mathsf{s}(t))))\\
\mathsf{o}(t)=&\rho'(\mathsf{s}(t))=\rho(f^{-1}(\mathsf{s}(t)))
\end{split}
\end{equation*}  
where the bijective function $f$ is shown in Fig.~\ref{fig:7t}. Therefore, the \BCN\ $\BB$ is identifiable.
\label{exa:103}
\end{example}




With {\em Lemma \ref{lemm:5}}, we can easily obtain the relationship between idenfiability and other properties. 
The controllability together with {\sf Type-III} observability is sufficient but not necessary for identifiability. The controllability together with {\sf Type-II} reconstructibility is neither sufficient nor necessary for identifiability, because the online observability does not imply the {\sf Type-II} reconstructibility and vice versa (it can be easily proved by listing the counterexamples).


%% file: sec/conclusions.tex
\section{Conclusions}
\label{sec:con}
In this paper, we formally defined the online observability, 
 proposed the algorithm to decide the online observability for \BCNs, and proved the relationship between identifiability and online observability.  


 But the decision algorithm of online observability mentioned in this paper has not been analyzed for complexity. Thus, we will research this problems and try to use some knowledge about formal methods to earn scalability for the online observability problem of \BCNs\ in our future work.



%% file: sec/appendix.tex
\section{Proof}
\label{sec:pro}

\subsection{Proof of Lemma \ref{lemm:5}}

\begin{pf}
(Sfficiency) First, if a \BCN\ $\BB$ is online observable, then we can construct a determining tree for it from its updating rules. And then, we have if a \BCN\ $\BB$ is controllable and online observable, we can construct a none state determining tree from its input-output data. 

Second, if we can construct a none state determining tree by $2^m$ ($m$ is the number of state-nodes) pairs of input sequences and output sequences from the input-output data of a \BCN\ $\BB$. Then, from the definition of none state determining tree, we have the $2^m$ sets of states which are determined by these $2^m$ pairs of input sequences and output sequences, respectively, are disjoint. As the total number of states is $2^m$, and for every set of there $2^m$ sets of states, it is not empty, so we have its cardinal number is $1$. Thus we can identify and determine all of the $2^m$ states by the none state determining tree.

Finally, if a \BCN\ $\BB$ is controllable and online observable, then we construct enough input-output data $\mathsf{i}(0)\ldots \mathsf{i}(t_0)$ and $\mathsf{o}(0)\ldots \mathsf{o}(t_0)$ which  satisfies that
\begin{itemize}
\item we can construct a none state determining tree by $2^m$ pairs of input sequences and output sequences from the input-output data. Then, we can regard the $2^m$ states which are determined by these $2^m$ pairs of input sequences and output sequences as the states $\mathsf{s}^0,\ldots,\mathsf{s}^{2^m -1}$ of a \BCN\ $\BB'$ (which is equivalent to $\BB$), respectively. Combining with the input-output data, we can construct the $\rho'$ for $\BB'$. 
\item And, in the input-output data, for every two distinct states $\mathsf{s}$ and $\mathsf{s}'$ of $\BB'$, there exists an input sequence which can make $\BB'$ reach  $\mathsf{s}'$ from $\mathsf{s}$. Then, for every $\mathsf{s}$ we can find an input sequence which can make $\BB'$ reach $\mathsf{s}$, i.e. $\mathsf{s}(t)=\mathsf{s}$. And as for every $\mathsf{s}(t)$, for every $\mathsf{i}(t)$, the $\mathsf{s}(t+1)=\sigma' (\mathsf{i}(t),\mathsf{s}(t))$ can be determined by the determining tree, we can construct the $\sigma'$ for the \BCN\ $\BB'$.
\end{itemize}

(Necessity) If a  \BCN\ $\BB$ does not satisfy the controllability or online observability, then we can not construct enough input-output data $\mathsf{i}(0)\ldots \mathsf{i}(t_0)$ and $\mathsf{o}(0)\ldots \mathsf{o}(t_0)$ from the \BCN\ $\BB$ which satisfies that we can construct a none state determining tree from the input-output data. And then, we can neither identify all of the  $2^m$ states of any \BCN\ $\BB'$ (which is equivalent to $\BB$) nor construct the updating rules for $\BB'$. Thus, the \BCN\ $\BB$ does not satisfy the identifiability.

\end{pf}

\subsection{Proof of Lemma \ref{lemm:4}}

\begin{pf}
Firsrly, we prove the propostion that for a set of possible state $\mathsf{S}(t)$, if there exists an input sequence $\mathsf{I}[t,t_k]\in\mathcal{I}^{[t,t_k]}$ for some $t_k >t$, such that for any distinct states $\mathsf{s}(t)$, $\mathsf{s}'(t) \in \mathsf{S}(t)$, $H^{[t,t_k]}(\mathsf{s}'(t),\mathsf{I}[t,t_k])\neq H^{[t,t_k]}(\mathsf{s}(t), \mathsf{I}[t,t_k])$, then $\Ks(\mathsf{S}(t))\ne\infty$.

\begin{itemize}
\item When $t_k=t+1$, for any two distinct states $\mathsf{s}(t)$, $\mathsf{s}'(t) \in \mathsf{S}(t)$, $H^{[t,t_k]}(\mathsf{s}'(t),\mathsf{I}[t,t_k])\neq H^{[t,t_k]}(\mathsf{s}(t), \mathsf{I}[t,t_k])$. Then we have for $\mathsf{S}(t)$,
 the input $\mathsf{i}(t)=\mathsf{I}[t,t_k]$ satisfies that
 \begin{itemize}
 \item  $|\Ded\left(\mathsf{S}(t),\mathsf{i}(t),\varepsilon\right)|=|\mathsf{S}(t)|$, and 
 \item  for every non-empty $\Ded(\mathsf{S}(t),\mathsf{i}(t),\mathsf{o}(t+1))$, the $\Ks(\Ded(\mathsf{S}(t),\mathsf{i}(t),\mathsf{o}(t+1)))=0$.
 \end{itemize}
Thus the $\Ks(\mathsf{S}(t))=1$, then the propostion is right when $t_k =t+1$.

\item If for $k=(t+1),\ldots, (t+n)$ the propostion is right. Then when $t_k=t+n+1$, for any distinct states $\mathsf{s}(t)$, $\mathsf{s}'(t) \in \mathsf{S}(t)$, $H^{[t,t_k]}(\mathsf{s}'(t),\mathsf{I}[t,t_k])\neq H^{[t,t_k]}(\mathsf{s}(t), \mathsf{I}[t,t_k])$. Then we have for $\mathsf{S}(t)$,
 there exists an input $\mathsf{i}(t)$ which is the first input of $\mathsf{I}[t,t_k]$, such that
 \begin{itemize}
\item  $|\Ded\left(\mathsf{S}(t),\mathsf{i}(t),\varepsilon\right)|=|\mathsf{S}(t)|$, and 
 \item  for every non-empty $\Ded(\mathsf{S}(t),\mathsf{i}(t),\mathsf{o}(t+1))$,\\ $\Ks(\Ded(\mathsf{S}(t),\mathsf{i}(t),\mathsf{o}(t+1)))\ne \infty$.
 \end{itemize}
Thus the $\Ks(\mathsf{S}(t))\ne \infty$, then the propostion is right when $t_k =t+n+1$.

As the propostion is right when $t_k =t+1$, and we have if for $t_k=(t+1),\ldots, (t+n)$ the propostion is right, then the propostion is right when $t_k=t+n+1$. Thus the propostion is right for any $t_k>t$.

Secondly, we have that if a \BCN\ satisfies the {\sf Type-III} observability, then there exists an input sequence $\mathsf{I}[t]\in\mathcal{I}^{[t]}$ for some $t>0$, such that for any two distinct states $\mathsf{s}(0)$, $\mathsf{s}'(0) \in \mathcal{S}$, $H^{[0,t]}(\mathsf{s}'(0),\mathsf{I}[t])\neq H^{[0,t]}(\mathsf{s}(0), \mathsf{I}[t])$.

Therefore, we have the for every non-empty \\$\Ded(\mathcal{S},\varepsilon, \mathsf{o})$, $\Ks(\Ded(\mathcal{S},\varepsilon,\mathsf{o}))\ne \infty$, and then the \BCN\ is online observable.
 \end{itemize}
\end{pf}

\subsection{Proof of Lemma \ref{lemm:3}}
\begin{pf} Firsrly, we prove the propostion that for a set of possible state $\mathsf{S}(t)$ if $\Ks(\mathsf{S}(t))\ne\infty$, then for every state $\mathsf{s}(t)\in \mathsf{S}(t)$, there exists an input sequence $\mathsf{I}[t,t_k]\in\mathcal{I}^{[t,t_k]}$ for some $t_k >t$ such that for every $\mathsf{s}'(t)\in \mathsf{S}(t)$, $\mathsf{s}'(t)\neq \mathsf{s}(t)$, $H^{[t,t_k]}(\mathsf{s}'(t),\mathsf{I}[t,t_k])\neq H^{[t,t_k]}(\mathsf{s}(t), \mathsf{I}[t,t_k])$.
\begin{itemize}
\item When $\Ks(\mathsf{S}(t))=0$, we have $|\mathsf{S}(t)|=1$, then for every $\mathsf{s}(t)$$\in \mathsf{S}(t)$ there does not exist any $\mathsf{s}'(t)\in \mathsf{S}(t)$ that $\mathsf{s}(t)\neq \mathsf{s}(t)$. Therefore, we have that for any $t_k >t$, for every input sequence $\mathsf{I}[t,t_k]\in\mathcal{I}^{[t,t_k]}$ the propostion is right. 
\item If for $\Ks(\mathsf{S}(t))=0,\ldots, n$ the propostion is right. When $\Ks(\mathsf{S}(t))=n+1$, we have for $\mathsf{S}(t)$ there exists a $\mathsf{i}(t)\in \mathcal{I}$ such that
 \begin{itemize}
 \item  $|\Ded\left(\mathsf{S}(t),\mathsf{i}(t),\varepsilon\right)|=|\mathsf{S}(t)|$, and 
 \item  for every non-empty $\Ded(\mathsf{S}(t),\mathsf{i}(t),\mathsf{o}(t+1))$,\\
  $\Ks(\Ded(\mathsf{S}(t),\mathsf{i}(t),\mathsf{o}(t+1)))<(n+1)$.
 \end{itemize}
 Then we have for every $\mathsf{s}(t)$$\in \mathsf{S}(t)$, there exists an $\mathsf{I}[t,t_k]\in\mathcal{I}^{[t,t_k]}$ for some $t_k >t$, such that for all $\mathsf{s}'(t)\in \mathsf{S}(t)$, $\mathsf{s}'(t)\neq \mathsf{s}(t)$ implies $H^{[t,t_k]}(\mathsf{s}'(t),\mathsf{I}[t,t_k])\neq H^{[t,t_k]}(\mathsf{s}(t), \mathsf{I}[t,t_k])$, where the input sequence $\mathsf{I}[k]=\mathsf{i}(t)\mathsf{I}[t+1,t_k]$. That the input sequence $\mathsf{I}[t+1,t_k]$ satisfies 
  $H^{[t+1,t_k]}(\mathsf{s}'(t+1),\mathsf{I}[t+1,t_k])\neq H^{[t+1,t_k]}(\mathsf{s}(t+1), \mathsf{I}[t+1,t_k])$, where $\mathsf{s}(t+1)=\sigma(\mathsf{s}(t),\mathsf{i}(t))$ and $\mathsf{s}'(t+1)=\sigma(\mathsf{s}'(t),\mathsf{i}(t))$. Then we have the propostion is right when $\Ks(\mathsf{S}(t))=n+1$. 

\end{itemize}
As the propostion is right when $\Ks(\mathsf{S}(t))=0$, and we have if for $\Ks(\mathsf{S}(t))=0,\ldots, n$ the propostion is right, then the propostion is right when $\Ks(\mathsf{S}(t))=n+1$. Thus the propostion is right for every $\Ks(\mathsf{S}(t))\ne\infty$.

Secondly, we have that if a \BCN\ is online observable,
then for every  non-empty $\Ded\left(\mathcal{S},\varepsilon, \mathsf{o}\right)$, $\Ks(\Ded\left(\mathcal{S},\varepsilon, \mathsf{o}\right))\ne \infty$

Therefore, we have for every initial state $\mathsf{s}(0)$$\in \mathcal{S}$, there exists an input sequence  $\mathsf{I}[t]\in\mathcal{I}^{[t]}$ for some $t>0$, such that for all states $\mathsf{s}'(0)\neq \mathsf{s}(0)$, $H^{[0,t]}(\mathsf{s}'(0),\mathsf{I}[t])\neq H^{[0,t]}(\mathsf{s}(0), \mathsf{I}[t])$. Thus the \BCN\ satisfies the  {\sf Type-I} observability if it is online observable.
\end{pf}

\subsection{Proof of Lemma \ref{lemm:1}}
\begin{pf}
If $\Ks(\mathsf{S}^{2})=0$, we have  \[\mathsf{S}^{1} = \mathsf{S}^{2}.\] Therefore, we have $\Ks(\mathsf{S}^{1})=\Ks(\mathsf{S}^{2})=0\ne\infty$, thus we have the {\em Lemma \ref{lemm:1}} is right when $\Ks(\mathsf{S}^{2})=0$.
 
 If for $\Ks(\mathsf{S}^{2})=0,\ldots, k$ the {\em Lemma \ref{lemm:1}} is right. When $\Ks(\mathsf{S}^{2})=k+1$, we have for $\mathsf{S}^{2}$ there exists a $\mathsf{i}\in \mathcal{I}$ such that
 \begin{itemize}
 \item  $|\Ded\left(\mathsf{S}^{2},\mathsf{i},\varepsilon\right)|=|\mathsf{S}^{2}|$, and 
 \item  $\Ks(\Ded(\mathsf{S}^{2},\mathsf{i},\mathsf{o}))<(k+1)$, for each non-empty $\Ded(\mathsf{S}^{2},\mathsf{i},\mathsf{o})$.
 \end{itemize}
 and there is not any $\mathsf{i} \in \mathcal{I}$ satisfies that
  \begin{itemize}
 \item  $|\Ded\left(\mathsf{S}^{2},\mathsf{i},\varepsilon\right)|=|\mathsf{S}^2|$,
 \item  $\Ks(\Ded\left(\mathsf{S}^{2},\mathsf{i},\mathsf{o}\right))< k$ for each non-empty $\Ded\left(\mathsf{S}^{2},\mathsf{i},\mathsf{o}\right)$.
 \end{itemize} 
 Then we have for $\mathsf{S}^{1}$ there exists a $\mathsf{i}\in \mathcal{I}$ such that
 \begin{itemize}
 \item  $|\Ded\left(\mathsf{S}^{1},\mathsf{i},\varepsilon\right)|=|\mathsf{S}^{1}|$, and 
 \item  $\Ks(\Ded(\mathsf{S}^{1},\mathsf{i},\mathsf{o}))<(k'+1)\le (k+1)$, for each non-empty $\Ded(\mathsf{S}^{1},\mathsf{i},\mathsf{o})\subseteq \Ded(\mathsf{S}^{2},\mathsf{i},\mathsf{o})$.
 \end{itemize}
 and there is not any $\mathsf{i} \in \mathcal{I}$ satisfies that
  \begin{itemize}
 \item  $|\Ded\left(\mathsf{S}^{1},\mathsf{i},\varepsilon\right)|=|\mathsf{S}^1|$,
 \item  $\Ks(\Ded\left(\mathsf{S}^{1},\mathsf{i},\mathsf{o}\right))<k'\le k$ for each non-empty $\Ded\left(\mathsf{S}^{1},\mathsf{i},\mathsf{o}\right)$, $\Ded\left(\mathsf{S}^{1},\mathsf{i},\mathsf{o}\right)\subseteq \Ded\left(\mathsf{S}^{2},\mathsf{i},\mathsf{o}\right)$,
 \end{itemize} 
 Therefore $\Ks(\mathsf{S}^{1})\ne\infty$, and we have if for $\Ks(\mathsf{S}^{2})=0,\ldots, k$ the {\em Lemma \ref{lemm:1}} is right, then when $\Ks(\mathsf{S}^{2})=k+1$ the {\em Lemma \ref{lemm:1}} is right too. 
And as the {\em Lemma \ref{lemm:1}} is right when $\Ks(\mathsf{S}^{2})=0$, we have the {\em Lemma \ref{lemm:1}} is right for every $\Ks(\mathsf{S}^{2})\ne \infty$.
 
\end{pf}
\subsection{Proof of Lemma \ref{lemm:2}}
\begin{pf}
For the non-empty $\mathsf{S}^{1}$ and $\mathsf{S}^{2}$, if $\mathsf{S}^{1}\subseteq\mathsf{S}^{2}$ and $\Ks(\mathsf{S}^{2})\ne\infty$.
For every $\mathsf{i}\in \Ri(\mathsf{S}^2)$, we have  $|\Ded\left(\mathsf{S}^2,\mathsf{i},\varepsilon\right)|=|\mathsf{S}^2|$, and 
for every $\mathsf{o} \in \mathcal{O}$, $\Ded(\mathsf{S}^2,\mathsf{i},\mathsf{o})\ne \emptyset$ implies $\Ks(\Ded(\mathsf{S}^2,\mathsf{i},\mathsf{o}))\ne \infty$. With {\em Lemma \ref{lemm:1}} we have $|\Ded\left(\mathsf{S}^1,\mathsf{i},\varepsilon\right)|=|\mathsf{S}^1|$, and 
for every $\mathsf{o} \in \mathcal{O}$, $\Ded(\mathsf{S}^1,\mathsf{i},\mathsf{o})\ne \emptyset$ implies $\Ks(\Ded(\mathsf{S}^1,\mathsf{i},\mathsf{o}))\le \Ks(\Ded(\mathsf{S}^2,\mathsf{i},\mathsf{o}))$, thus $\mathsf{i}\in \Ri(\mathsf{S}^1)$. Therefore, $\Ri(\mathsf{S}^2)\subseteq\Ri(\mathsf{S}^1)$.
\end{pf}

\section{Algorithm}
\label{sec:algo}
\begin{algorithm}[h]
\caption{Determination algorithm}
\begin{algorithmic}[1]
\Require The updating rules of a \BCN
\Ensure  The input-labelled graph of this \BCN
\State Boolean value $Ob=$ true 
\State integer $i$, $z=1$\
\State array $VertexArray[\ ]$, $InputArray[\ ]$
\State {\sf constructvertex}($z$)
\State $VertexArray=${\sf constructvertex}($++z$)
\While {($VertexArray!=$Null)}
\For{($i=0$; $i<arraysize(VertexArray)$; $i++$)}
\If{($z==2$)}
\State $InputArray$ = $\mathcal{I}_\BB$ 
\Else
\State $InputArray$=$\mathop{\bigcap}\limits_{\mathsf{S}\subset VertexArray[i]} \Ri(\mathsf{S}) $ 
\If{($InputArray==\emptyset$)}
\State  $Ob=$ false 
\State Return Null
\EndIf
\EndIf
\State Get $\Ri(VertexArray[i])$ by $InputArray$ 
\If {($\Ri(VertexArray[i])==\emptyset$)}
\State  $Ob=$ false 
\State Return Null
\EndIf
\State Build edges for $VertexArray[i]$ 
\EndFor
\State $VertexArray=${\sf constructvertex}($++z$)
\EndWhile
\State Return {\sf constructvertex}($--z$)
\end{algorithmic}
 \label{alg:1}
\end{algorithm}

\begin{algorithm}[h!]
\caption{{\sf constructvertex}(integer $z$)}
\label{alg:2}
\begin{algorithmic}[1]
\Require The number of states $z$
\Ensure  The vertexes with $z$ states producing the same outputs 
\State array $VertexArray[\ ]$
\State $VertexArray=\{\mathsf{S}\in\bigcup\limits_{\mathsf{o}\in\mathcal{O}} 2^{\Ded(\mathcal{S}_\BB,\varepsilon,\mathsf{o} )}|\ |\mathsf{S}|=z\}$
\If{($VertexArray==\emptyset$)} 
\State  Return Null
\Else 
\State  Classify these vertexes
\State Sort the states in these vertexes
\State Sort these vertexes
\State Return $VertexArray$
\EndIf 
\end{algorithmic}
\end{algorithm}